\documentclass[11pt]{article}
\usepackage{jheppub}

\usepackage{epsfig}
\usepackage{amssymb}
\usepackage{amsfonts}
\usepackage{amsbsy}
\usepackage{amsmath}
\usepackage{dsfont}
\usepackage{bbm}
\usepackage{upgreek}
\usepackage{amssymb,amscd}
\usepackage{graphicx}
\usepackage{mathrsfs}
\usepackage{amsmath,amsthm}
\usepackage{slashed}
\usepackage{slashbox}
\usepackage{dsfont}
\usepackage{tikz}
\usepackage[utf8]{inputenc}
\usepackage{tikz-feynman}\tikzfeynmanset{compat=1.0.0}

\usetikzlibrary{positioning}
\usetikzlibrary{shapes.geometric, arrows}
\tikzstyle{rect} = [rectangle,rounded corners,minimum width=3cm, minimum height=1cm, text centered, draw=black]
\tikzstyle{arrow} = [thick,->,>=stealth]
\usepackage{stackrel}
\tikzstyle{oct} = [regular polygon,regular polygon sides=8, draw,
    text width=2em, text centered]
\tikzstyle{line} = [draw, -latex']

\tikzstyle{hex} = [regular polygon,regular polygon sides=6, draw,
    text width=2em, text centered]
\tikzstyle{hexs}= [regular polygon,regular polygon sides=5, draw,
    text width=1.7em, text centered]

\makeatletter
\DeclareFontFamily{OMX}{MnSymbolE}{}
\DeclareSymbolFont{MnLargeSymbols}{OMX}{MnSymbolE}{m}{n}
\SetSymbolFont{MnLargeSymbols}{bold}{OMX}{MnSymbolE}{b}{n}
\DeclareFontShape{OMX}{MnSymbolE}{m}{n}{
    <-6>  MnSymbolE5
   <6-7>  MnSymbolE6
   <7-8>  MnSymbolE7
   <8-9>  MnSymbolE8
   <9-10> MnSymbolE9
  <10-12> MnSymbolE10
  <12->   MnSymbolE12
}{}
\DeclareFontShape{OMX}{MnSymbolE}{b}{n}{
    <-6>  MnSymbolE-Bold5
   <6-7>  MnSymbolE-Bold6
   <7-8>  MnSymbolE-Bold7
   <8-9>  MnSymbolE-Bold8
   <9-10> MnSymbolE-Bold9
  <10-12> MnSymbolE-Bold10
  <12->   MnSymbolE-Bold12
}{}

\let\llangle\@undefined
\let\rrangle\@undefined
\DeclareMathDelimiter{\llangle}{\mathopen}%
                     {MnLargeSymbols}{'164}{MnLargeSymbols}{'164}
\DeclareMathDelimiter{\rrangle}{\mathclose}%
                     {MnLargeSymbols}{'171}{MnLargeSymbols}{'171}
\makeatother

\def\be{ \begin{equation} }
\def\ee{ \end{equation}}

\newcommand{\eq}[1]{\begin{align}\begin{split}#1\end{split}\end{align}}

\def\vr{{\vec{r}}}

\def\cot{{\rm cot}}

\def\dim{{\rm dim}}
\def\exp{{\rm exp}}

\def\ker{{\rm ker}}

\def\sign{{\rm sign}}

\def\half{\frac{1}{2}}

\def\ihalf{\frac{i}{2}}

\def\g{\frac{1}{g^2}}

\def\one{{\hbox{ 1\kern-.8mm l}}}

\def\vT{{\vec{T}}}
\def\vI{{\vec{I}}}
\def\vJ{{\vec{J}}}
\def\vK{{\vec{K}}}
\def\vS{{\vec{S}}}
\def\vL{{\vec{L}}}
\def\vA{{\vec{A}}}

\def\CD {{\cal D}}

\def\CK {{\cal K}}

\def\CM {{\cal M}}
\def\CN {{\cal N}}

\def\CP {{\cal P}}

\def\CY {{\cal Y}}

\def\IC{\mathbb{C}}

\def\IR{{\mathbb{R}}}

\def\IZ{{\mathbb{Z}}}

\def\fg{\mathfrak{g}}

\def\fs{\mathfrak{s}}

\def\fsu{\mathfrak{su}}

\def\fu{\mathfrak{u}}

\def\fM{\mathfrak{M}}

\def\thm#1{\bigskip\noindent{\bf Theorem #1} }
\def\corro#1{\bigskip\noindent{\bf Corollary #1} }
\def\lem#1{\bigskip\noindent{\bf Lemma #1} }

\def\cnj#1{\bigskip\noindent{\bf Conjecture:} }
\def\pf{\noindent{\textit{proof.}} }

\DeclareMathAlphabet{\mathpzc}{OT1}{pzc}{m}{it}

\def\diag{{{\rm diag}}}

\title{Index-like Theorem for Massless Fermions in Spherically Symmetric Monopole Backgrounds
}

\author[a]{T.~Daniel Brennan}

\affiliation[a]{Kadanoff Center for Theoretical Physics \& Enrico Fermi Institute, \\
University of Chicago,
Michelson Center for Physics, 933 E 56th St, Chicago, IL 60637. 
}

\emailAdd{tdbrennan@uchicago.edu}

\abstract{ 
In this paper we study massless fermions coupled to spherically symmetric $SU(N)$ monopoles without Yukawa couplings between the Higgs and fermion fields. The corresponding Dirac operator is not Fredholm and the associated eigenfunctions are not $L^2$-normalizable. Here  we derive a formula for the dimension of the plane-wave normalizable kernel of such a Dirac operator for fermions of any representation of $SU(N)$ in the presence of any spherically symmetric monopole background. Notably, our results also apply to fermions coupled to monopoles that preserve non-abelian gauge symmetry. 
}

\begin{document}

\maketitle

\section{Introduction and Summary}

In this paper we are concerned with interactions of monopoles with fermions in 4D $SU(N)$ gauge theories. An important feature of  these interactions is the spectrum of fermion zero-modes. Fermions admit $L^2$-normalizable zero-modes when  the spectrum of eigenfunctions of the Dirac operator  has an energy gap which typically comes from either a mass  term or from coupling to the Higgs field via a Yukawa interaction. 
This energy gap implies that the Dirac operator is Fredholm and the number of fermion zero-modes is given by its index \cite{Callias:1977kg}. 
Let  us denote this energy gap $E_{\rm gap}$. 
In the case of $E_{\rm gap}>0$, the fermion zero-modes decay exponentially away from the monopole as 
\eq{
\lim_{r\to \infty}\psi(r,\theta,\phi)\sim \frac{1}{r}e^{- |E_{\rm gap}| r}\psi_0(\theta,\phi)~,}
confining the zero-modes to the world volume of the monopole. 
The low energy dynamics of the fermions then reduces to the quantum mechanics of these zero-modes which  can impart the monopole with charges under global symmetries and can affect monopole-monopole interactions \cite{Jackiw:1975fn,Gauntlett:1993sh,Sethi:1995zm,Gauntlett:1999vc,Gauntlett:2000ks,Brennan:2016znk,Brennan:2018ura}. 


Here, we would like to study the limit $E_{\rm gap}\to 0$ arising from sending the strength of the Yukawa interaction to zero. In this case, generically some collection of 
 zero-modes become non-$L^2$-normalizable.
In sending $E_{\rm gap}\to 0$, these zero-modes become no longer localized on the monopole world volume, but rather describe free propagating fields that can tunnel into the core of the monopole and interact with the non-abelian gauge degrees of freedom that are trapped there. This is known as the Callan-Rubakov effect and results in monopoles that can catalyze proton decay in grand unified theories such as the $SU(5)$ Georgi-Glashow model \cite{Georgi:1974sy,Callan:1982ah,Callan:1982au,Callan:1982ac,Rubakov:1982fp}.  

In this paper, we will be concerned with the case where all zero-modes become non-$L^2$-normalizable from decoupling the fermions frmo the Higgs field. 
In this situation, where $E_{\rm gap}\to 0$, the zero-energy solutions belong to a continuous spectrum of spherical scattering states that goes down to zero-energy. These scattering solutions 
fall into families based on the zero-energy solutions 
of the Dirac equation. 
 Thus,  the mathematically relevant question is to determine the dimension of the kernel of the Dirac operator when restricted to solutions that are ``plane-wave normalizable'':
\eq{
\lim_{r\to \infty}r\psi<\infty~,
}
and smooth. 


The problem of understanding the spectrum of non-Fredholm Dirac operators is also closely related to the phenomenon of wall crossing in supersymmetric gauge theories. In 4D $\CN=2$ gauge theories, the stability of BPS states can be determined by solving the Dirac equation on monopole moduli space ($\CM$) coupled to a Killing vector field \cite{Gauntlett:1993sh,Gauntlett:1999vc,Gauntlett:2000ks,Brennan:2016znk,Brennan:2018ura}. Approaching a wall of marginal stability corresponds to shifting the Killing vector so that the Dirac operator on $\CM$ becomes non-Fredholm and the solutions to the Dirac equation are no longer $L^2$-normalizable. This  is analogous to solving the Dirac equation in the presence of a monopole 
with a Yukawa coupling $\lambda_{Yuk}$ in the limit $\lambda_{Yuk}\to 0$  (sends $E_{\rm gap}\to 0$). In fact, in the asymptotic limit of moduli space near a wall of marginal stability where the BPS states separate into two relatively unstable clusters, 
the wall crossing is controlled by a 4D Dirac operator coupled to an abelian monopole that becomes non-Fredholm on the wall of marginal stability. In this case, extra data is needed to understand the regular solutions of the Dirac operator in the limit $r\to 0$. 

\subsection{Summary and Conclusion}

In this paper, we derive a formula to enumerate the number of plane-wave normalizable zero-energy solutions of the Dirac equation in the presence of a spherically symmetric monopole. 

The setting for this paper is four-dimensional $SU(N)$ gauge theory with gauge field $A_\mu$ coupled to a real, adjoint Higgs field $\Phi$ and a  Weyl fermion $\psi_R$ that transforms under a representation $R$ of $SU(N)$. Here we will only consider the theory with no Yukawa coupling between $\Phi,\psi_R$.

We would like to consider the time-independent Dirac equation for the fermion 
\eq{
i \bar\sigma^\mu D_\mu \psi_R=0~,
}
where $D_\mu=\partial_\mu +i R(A_\mu)$ and $\mu=1,2,3$ 
in the presence of a spherically symmetric BPS monopole. 

A (BPS) monopole is  a Yang-Mills-Higgs configuration that satisfies the Bogomolny equation 
\eq{
B_\mu=D_\mu \Phi~,
}
with the asymptotic behavior
\eq{
\lim_{r\to \infty}\vec{B}&=\frac{\gamma_m}{r^2}\hat{r}+O(1/r^{2+\delta})\quad, \quad 
\lim_{r\to \infty}\Phi=\Phi_\infty-\frac{\gamma_m}{r}+O(1/r^{1+\delta})\quad\delta>0~,
}
where $\Phi_\infty\in \fsu(N)$ is the asymptotic Higgs field and $\gamma_m\in \fsu(N)$ is the asymptotic magnetic charge which satisfies $[\gamma_m,\Phi_\infty]=0$. Here $\Phi_\infty$ breaks $SU(N)$ to the centralizer $G_{\Phi_\infty}=\{g\in SU(N)~\big{|}~g^{-1}\,\Phi_\infty \,g=\Phi_\infty\}$. Here we will \emph{not} assume that $G_{\Phi_\infty}$ is an abelian group. 

A spherically symmetric monopole is a special class of monopole that is rotationally invariant under $\vK=-i \vec{r}\times \vec\nabla+\vT$ where $\{T_i\}_{i=1}^3\in \fsu(N)$ (which satisfy $[T_i,T_j]=i\epsilon_{ijk}T_k$)  generate a $SU(2)_T$ subgroup of the gauge group $SU(N)$  \cite{Wilkinson:1977yq}. Such a gauge field configuration can be specified by an additional choice of embedding $SU(2)_I\hookrightarrow SU(N)$ with generators 
\eq{
I_i\in \fs\fu(N)\quad,  \quad [I_i,I_j]=i\epsilon_{ijk}I_k \quad, \quad i=1,2,3~,
}
such that 
\eq{
\gamma_m=T_3-I_3\quad, \quad 
[\gamma_m,I_i]=0~.
}
We will further require that given a decomposition of $\vT$ into irreducible components
\eq{
\vT=\bigoplus_i \vT^{(i)}~,
}
that we can also decompose $\vI$ into irreducible components satisfying
\eq{
\vI=\bigoplus_{i,a} \vI^{\,(i,a)}\quad, \quad [\vT^{(i)},\vI^{\,(j,a)}]=0~,~i\neq j~,
}
and we assume $\vI^{(i,a)}\neq \vT^{(i)}$ for any $i,a$. 
%
This data specifies the vacuum expectation value of the Higgs field 
\eq{
\Phi_\infty=\sum_{i}v_i\left(T_3^{(i)}-\sum_a I_3^{(i,a)}\right)+\phi_\infty~,
}
up to a choice of $v_i\in \IR_{\geq0}$ and $\phi_\infty \in \fsu(N)$ such that $[\phi_\infty,\vT^{(i)}]=[\phi_\infty,\vI^{\,(i,a)}]=0$, $\forall i,a$. 

Now we wish to find the time-independent solutions to the Dirac equation that are smooth, bounded, and ``plane-wave normalizable'':
\eq{
\lim_{r\to \infty} r\,\psi_R(r,\theta,\phi)&<\infty~. 
}
Here we will prove that the number of such zero-energy solutions of the Dirac equation is
\eq{\label{dimeq}
\dim_\IC\big[\ker_R[i\bar\sigma^\mu D_\mu]\big]&=2\sum_{\mu \in \Delta^+_R[T_3,I_3]}n_R(\mu)\,\langle \mu,T_3\rangle
\\
&=\lim_{\epsilon\to 0^+} \sum_{\mu \in \Delta_R}n_R(\mu)\langle \mu,T_3\rangle\, \sign\left(\langle \mu-\epsilon w(\mu), T_3- I_3\rangle\right)~,
}
where $\Delta_R\subset \Lambda_{wt}(\fsu(N))$ is the weight system of the representation $R$,  $n_R(\mu)$ is the multiplicity of $\mu\in \Delta_{R}$, and 
\eq{
\Delta^+_{R}[T_3,I_3]=\left\{\mu \in \Delta_{{R}}^+~\big{|}~\langle \mu,T_3-I_3\rangle\langle w(\mu),T_3-I_3\rangle<0\right\}~,
}
 where $w\in W(SU(N))$ is the generator of the Weyl group of $\Lambda_{rt}[\fsu(2)_T]\subset \Lambda_{rt}[\fsu(N)]$ and $\Delta_R^+$ is the set of positive weights of $R$ with respect to $T_3\in \fsu(N)$. 
 Further, we can show that as representations of the total angular momentum $\vJ=\vL+\vS+\vT$, the kernel decomposes as 
\eq{\label{kereq}
\ker_R[i\bar\sigma^\mu D_\mu]=\bigoplus_{\mu \in \Delta^+_R[T_3,I_3]}\left[\langle \mu,T_3\rangle-\half\right]^{\oplus n_R(\mu)}~,
}
where $[j]$ is the spin-$j$ representation.

 This work is directly applicable to understanding the Callan-Rubakov effect in which low energy, massless fermions scatter off of a monopole in a way that is sensitive to the UV physics inside the monopole. 
 If we consider massless fermions propagating in the presence of a spherically symmetric monopole, then the low energy dynamics of the fermions are governed by 2D massless fermions that correspond to 
radially propagating waves 
 in the presence of the asymptotic, abelian monopole field. The Callan-Rubakov effect 
 can then be attributed to the effective boundary conditions 
 imposed on the fermions at the monopole core, which can be read off from the form of the zero-energy fermion solutions of the full non-abelian theory found in this paper. This story will be addressed in further detail in an separate paper \cite{Brennan:2021ewu}. 


The outline of this paper is as follows. In section \ref{sec:2} we will discuss spherically symmetric monopoles. Then in section \ref{sec:3} we will explicitly solve the Dirac equation in the spherically symmetric monopole background and show that imposing plane-wave normalizability is equivalent to a  representation theory problem. In section \ref{sec:4} we will derive the main results \eqref{dimeq} and \eqref{kereq} which we then illustrate with an example in section \ref{sec:5}.  

\section{Spherically Symmetric Monopoles}
\label{sec:2}


In this paper, we will consider a class of BPS monopoles. A BPS monopole in 4D $G$-gauge theory is a finite energy, time-independent gauge field configuration with vanishing electric field that solves the Bogomolny equation 
\eq{
B_\mu=D_\mu \Phi~,
}
where $B_\mu$ is the magnetic field. 
 The solutions of this equation are separated into distinct classes by the asymptotic behavior
\eq{
\lim_{r\to \infty}\vec{B}= \frac{\gamma_m}{r^2}\hat{r}+O(1/r^{2+\delta})\quad, \quad \lim_{r\to \infty}\Phi=\Phi_\infty-\frac{\gamma_m}{r}+O(1/r^{1+\delta})\quad \delta>0~,
}
where $\Phi_\infty\in \fg$ is the vacuum expectation value of the Higgs field which breaks gauge symmetry $G\to G_{\Phi_\infty}$ to the centralizer subgroup and $\gamma_m\in \fg_{\Phi_\infty}$ is the asymptotic magnetic charge\footnote{
Here we make the choice $2\gamma_m \in \Lambda_{cr}(SU(N))$. We make this choice of convention to simplify formulas later. 
}.
%
In this paper we will  take $\Phi_\infty\neq 0$, but we will not assume that $G_{\Phi_\infty}$ is abelian. Additionally, when $G_{\Phi_\infty}$ we will generally consider monopoles that have vanishing holomorphic charge \cite{Murray:2003,Bais:1997qy}. 

With these asymptotics, it is clear that the gauge field can asymptotically be put into the form:\footnote{Note that here there is a difference of a factor of 2 from other standard references. This follows from our choice of convention that $\gamma_m$ obeys the quantization $e^{4\pi i \gamma_m}=\mathds{1}_{SU(N)}$ and  $\phi$ has periodicity $\phi\sim \phi+2\pi$ which additionally matches with our choice  of $2\gamma_m\in \Lambda_{cr}(SU(N))$. 
}
\eq{
A_{ab.}=2\gamma_mA_{Dirac}+O(1/r^2)=\gamma_m(\sigma -\cos(\theta))d\phi+O(1/r^2)~,
}
where $\sigma=\pm1$ for in the coordinate patch containing the $N/S$-hemisphere. Without loss of generality, for the remainder of the paper we will work in northern hemisphere where $\sigma=1$.

A large class of smooth monopoles can be constructed by embedding the 't Hooft-Polyakov $SU(2)$ monopole into the gauge group $G$ \cite{tHooft:1974kcl,Polyakov:1974ek,Prasad:1975kr,Weinberg:2006rq}.  
However, not all monopoles can be put in this form. 
More generally, we can consider spherically symmetric monopoles which are not simple embeddings of the $SU(2)$ monopole. 
Let us recall the construction of Wilkinson-Bais \cite{Wilkinson:1978zh} for higher charge $SU(N)$ monopoles. 

 Since a monopole is a gauge field configuration and not simply a vector field, a monopole is spherically symmetric if it is rotationally invariant up to a gauge transformation. 
To define a spherically symmetric gauge field configuration, we can augment the standard angular momentum generators by the generators of an $SU(2)$ subgroup of the gauge group \cite{Wilkinson:1977yq,Wilkinson:1978zh} which we will denote $SU(2)_T$:
\eq{
\vec{K}=\vec{L}+\vec{T}=-i \vec{r}\times\vec{\nabla}
+ \vec{T}~,
} 
where $T_i\in \fsu(N)$  satisfy $[T_i,T_j]=i \epsilon_{ijk}T_k$. It is clear that the generators $\{K_i\}_{i=1}^3$ generate a $\fsu(2)_K$ algebra. In this paper, we will restrict to the set of spherically symmetric monopoles that are symmetric with respect to this modified rotation generator. 

For a field configuration to be spherically symmetric, we require that 
\eq{
[K_i, A_j]=i \epsilon_{ijk} A_k~.
}
Let us take the ansatz
\eq{\label{vecgauge}
\vec{A}=\frac{\hat{r}}{r}\times (\vec{\fM}-\vec{T})~. 
}
 We will refer to this form as the \emph{vector gauge}.

Our ansatz defines a spherically symmetric gauge solution (satisfying $[K_i,A_j]=i \epsilon_{ijk}A_k$)  if 
\eq{
[K_i,\fM_j]=i \epsilon_{ijk}\fM_k~. 
}
Such a $\fM_i$ can be constructed by taking 
\eq{\fM_i= \Omega (\theta,\phi)\, M_i (r)\, \Omega ^{-1}(\theta,\phi)
\quad, \quad M_3(r)=0~,
}
where $\vec{M}=(M_1(r)\,,\,M_2(r)\,,\,0)$ 
such that 
\eq{
[T_3,M_\pm]=\pm M_\pm\quad, \quad M_\pm=M_1\pm i M_2~.
}
Here $ \Omega (\theta,\phi)$ is a gauge transformation that acts as a rotation matrix in $SU (2)_T$ that maps $\Omega^{-1}(\hat{r}\cdot\vec{T})\Omega= T_3$. This produces the rotationally invariant $\vec\fM$ because $\Omega(\theta,\phi)$ acts on $\vec{M}$ as the rotation matrix generated by $\vec{K}$. 

Using this, we see that  the gauge field is determined by its value along 
 the $\hat{z}$-axis:
\eq{
\vec{A}\big{|}_{z\text{-axis}}=\frac{\hat{r}}{r}\times \left(\vec{M}(r)-\vec{T}\right)~.
}

If we explicitly parametrize 
\eq{
M_+(r)&=\big(M_-(r)\big)^T=\left(\begin{array}{ccccc}
0&\alpha_1(r)&&&\\
&0&\alpha_2(r)&&\\
&&\ddots&\ddots&\\
&&&0&\alpha_{N-1}(r)\\
&&&&0
\end{array}\right)~,\\
\Phi\big{|}_{z\text{-axis}}&=\diag(\phi_1(r)\,,\,\phi_2(r)-\phi_1(r)\,,\,...\,,\,-\phi_{N-1}(r))~,
} 
then we find that  the Bogomolny equation reduces to the coupled differential equations
\eq{
\frac{d\phi_I(r)}{dr}&=\frac{1}{r^2}(\alpha_I^2(r)-I(N+1-I))~,\\
\frac{d\alpha_I(r)}{dr}&=\half(2\phi_I(r)-\phi_{I-1}(r)-\phi_{I+1}(r))\alpha_I(r)~.
}
These equations have explicitly known solutions as described in \cite{Wilkinson:1978zh} that are smooth, bounded and have finite energy. These solutions have an asymptotic magnetic charge   $\gamma_m\in \fs\fu(N)$ which is related to an additional embedding $SU(2)_{I}\hookrightarrow SU(N)$ 
 specified by $\{I_i\}\in \fs\fu(N)$ for $i=1,2,3$ by 
\eq{
\gamma_m=T_3-I_3~,
}
where we require that $[\vI,\gamma_m]=[\vI,\Phi_\infty]=0$. We also require that given a decomposition of  the $\vT,\vI$ into irreducible components:
\eq{
\vT=\bigoplus_i \vT^{(i)}\quad, \quad 
\vI=\bigoplus_{i,a} \vI^{\,(i,a)}~,
}
 that obey
\eq{
[\vT^{(i)},\vI^{\,(j,a)}]=0~,~i\neq j~,
}
and we impose $ \vI^{(i,a)}\neq \vT^{(i)}$ for any $i,a$. 

With this data, the asymptotic value of the Higgs field of the monopole configuration is given by 
\eq{
\Phi_\infty=\sum_i v_i\left(T_3^{(i)}-\sum_aI_3^{(i,a)}\right)+\phi_\infty~,
}
for some $v_i\in \IR$ and $\phi_\infty\in \fsu(N)$ such that $[\vT^{(i)},\phi_\infty]=[\vI^{\,(i,a)},\phi_\infty]=0$, $\forall a,i$. For simplicity, we will only consider $\phi_\infty=0$. In this case, the Higgs vev breaks $SU(N)$ to a subgroup with non-abelian gauge symmetry preserved along each non-trivial irreducible $\vI$ embedding.




For our purposes, it will be more convenient to write the connection in a different gauge which is related to the vector gauge by the transformation $g=e^{-i \phi T_3}e^{- i \theta T_2}e^{i \phi T_3}$:
\eq{
A_{can}=2 T_3A_{Dirac}+i\frac{M_+}{2}e^{-i \phi}(d\theta-i \sin\theta d\phi)-i \frac{M_-}{2} e^{i \phi}(d\theta+i \sin\theta d\phi)~.
}
We will refer to this form as the \emph{canonical gauge}. 

Using the fact that $[T_3,M_\pm]=\pm  M_\pm$, we can explicitly can compute the field strength:
\eq{
F=&dA+i[A,A]
=\left(T_3 -\frac{ [M_+,M_-]}{2} \right)d^2\Omega\pm  \ihalf M'_\pm(r) e^{\mp i \phi}dr\wedge (d\theta\mp i \sin\theta d\phi)~,\\
}
where the $\pm$ are correlated and summed over. In vector notation, the field strength 
reduces along the $\hat{z}$-axis to
\eq{\label{vecfield}
B_r=\frac{T_3}{r^2}-\frac{[M_+,M_-]}{2r^2}\quad, \quad B_\pm =\frac{1}{r}\partial_r M_\pm ~,
}
%
Note that the normalizability of the gauge field \eqref{vecgauge} and the asymptotic form of the field strength \eqref{vecfield} fixes 
\eq{
&\lim_{r\to 0}M_\pm =T_\pm+O(r)\quad, \quad \lim_{r\to \infty}M_\pm =I_\pm +O(1/r)~.
}

\section{Fermion Zero-Modes}
\label{sec:3}

Now we would like to solve for the zero energy solutions of the Dirac equation coupled to a spherically symmetric monopole.

\subsection{Fermions and Angular Momentum} 

\label{sec:fermrep}

As we have discussed, spherically symmetric monopole configurations transform as a vector under $\vK=\vL+ \vT$. This implies that the Dirac operator  commutes with \eq{
\vJ=\vL+\vS+\vT=-i \vec{r}\times \vec\nabla+\half \vec\sigma+\vT~,}
which generate a group $SU(2)_J$. 
 Therefore, the solutions of the Dirac equation can be decomposed in terms of representations of $SU(2)_J$.

Let us first note that $\vJ=\vL+\vS+\vT$ is the sum of three mutually commuting $\fs\fu(2)$ generators. Therefore, the representation of $SU(2)_J$ follows the standard addition of three $SU(2)$ representations. In particular, this means that the $SU(2)_J$-representations will be specified by a representation of $SU(2)_T$, a spin, and an orbital angular momentum. 

Thus, given a fermion $\psi_R$ that transforms under a representation $R$ of $SU(N)$, the $SU(2)_J$-representations will decompose with respect to the restriction of $R$ to irreducible $SU(2)_T\subset SU(N)$ representations:
\eq{
R=\bigoplus_A R_A\quad\Rightarrow \quad \psi_R=\sum_A\psi_{R_A}~,
}
where $R_A$ is a $SU(2)_T$ representation of spin-$q_{R_A}$ and $\psi_{R_A}$ transforms under $R_A$. 
The dimension of the $S$- and $T$- representations are fixed to $2$ and $2q_{R_A}+1$. In general, we find that there will be multiple spin-$j$ representations which come from the representations of $L$ (with quantum number $\ell$) such that 
\eq{
[j]\in [\ell]\otimes \left[\half\right]\otimes [q_{R_A}]\quad, \quad \ell\in \IZ_{\geq0}~,
}
where $[n]$ is the spin-$n$   representation. 
These multiple spin-$j$ representations will span the space of spin-$j$ representations of $SU(2)_J$.

Now when we gauge transform from the vector to the canonical gauge
\eq{
A_{can}&=  g ^{-1} A_{vec}  g -i  g ^{-1}d   g \\
&=2T_3 A_{Dirac}+i \frac{M_+}{2} e^{-i \phi}(d\theta-i \sin  \theta d\phi)-i \frac{M_-}{2}e^{i \phi}(d\theta+i\sin\theta d\phi)~, 
}
where $g=e^{-i \phi T_3}e^{- i \theta T_2}e^{i \phi T_3}$, 
the total angular momentum generators $\vJ$ also transform
%
:
\eq{
\vec{J}\to -i \vr\times \Big(\vec\partial+2i T_3 \vec{A}_{Dirac}\Big)+T_3\hat{r}+\half \vec\sigma~.
}
Additionally,  we will find it useful to implement a frame rotation by
\eq{
U_\alpha^{~\beta}(\theta,\phi)=e^{\frac{i\phi}{2}\sigma^0\bar\sigma^3}e^{\frac{i \theta}{2}\sigma^0\bar\sigma^2}~,
}
when solving the Dirac equation. 
This also transforms $\vJ$ into the form
\eq{
\vJ\to-i \vr\times \left(\vec\partial+2i T_3\vec{A}_{Dirac}-i\sigma^3\half \cot\theta\,\hat\phi\right)+\left(T_3+\frac{\sigma_3}{2}\right)\hat{r}~.
}
Now we would like to construct the representations of $SU(2)_J$ by diagonalizing $J^2,J_3$ which have explicit forms:
\eq{
J^2&=-\half\left[D^-_{T_3+\frac{\sigma^3}{2}+1}D^+_{T_3+\frac{\sigma^3}{2}}+D^+_{T_3+\frac{\sigma^3}{2}-1}D^-_{T_3+\frac{\sigma^3}{2}}\right]
+\left(\frac{\sigma_3}{2}+T_3\right)^2~,\\
J_3&=-i\big( \partial_\phi+iT_3\big)~,
}
where 
\eq{
D^\pm_q=\partial_\theta\mp \frac{1}{\sin\theta}\Big(i(\partial_\phi+i T_3)+q \cos\theta\Big)~.
}

Now consider a fermion that transforms under the $SU(2)_T$-irreducible representation $R_A$, $\psi_{R_A}$. 
Let us write the basis of the representation space 
\eq{
V_{R_A}={\rm span}_\IC\{v_{\mu_a}\}\quad, \quad \mu_a\in \Delta_{R_A}~,
}
where $\Delta_{R_A}$ is the weight space of $R_A$. 
For each $R_A$, we will pick a basis $\{\mu_a\}$ of $\Delta_{R_A}$
 which such that $\langle \mu_a,T_3\rangle>\langle \mu_{a+1},T_3\rangle$ and $\langle \mu_a,\gamma_m\rangle\geq\langle \mu_{a+1},\gamma_m\rangle$. Then, if we decompose 
\eq{
\psi_{R_A}=\sum_a \psi^a_{R_A} v_{\mu_a}\quad, \quad {R_A}(T_3)\cdot v_{\mu_a}=q_a v_{\mu_a}\quad, \quad q_a=\langle \mu_a,T_3\rangle~, 
}
we find that the eigenfunctions  of $J^2,J_3$ are given by 
\eq{
\psi^a_{R_A}
&=\sum_{s=\pm}f_{s,a}(r)\CY^j_{m,q_a,s}(\theta,\phi)\\
&=e^{i (m-q_a)\phi}\left(\begin{array}{c}f_{a,+}(r)\,\,d^{j}_{-m,q_a+\half}(\theta)\\
f_{a,-}(r)\,d^{j}_{-m,q_a-\half}(\theta)
\end{array}\right)
=\left(\begin{array}{c}f_{a,+}(r)\,e^{i \phi/2}\,\CD^{(j)}_{-m,q_a+\half}(\theta,\phi)\\
f_{a,-}(r)\,e^{-i \phi/2}\CD^{(j)}_{-m,q_a-\half}(\theta,\phi)
\end{array}\right)~,
}
 where
$d^j_{m,q}(\theta)$ are small Wigner $d$-functions and $\CD^{(j)}_{m,q}(\theta,\phi):=\CD^{(j)}_{m,q}(\phi,\theta,\phi)$ are Wigner $D$-functions. The above representation has eigenvalues 
\eq{
J^2\,\CY^j_{m,q_a,s}(\theta,\phi)=j(j+1)\CY^j_{m,q_a,s}(\theta,\phi)\quad, \quad J_3\, \CY^j_{m,q_a,s}(\theta,\phi)=m \CY^j_{m,q_a,s}(\theta,\phi)~.
}
This follows from the fact that 
\eq{
D^\pm_{q_a} \left[e^{i(m-q_a)\phi}d^j_{-m,q_a}(\theta)v_{\mu_a}\right]&=\mp \sqrt{j(j+1)-q_a(q_a\pm 1)}\left[e^{i(m-q_a)\phi}d^j_{-m,q_a\pm1}(\theta)v_{\mu_a}\right]~.
}

The form of these eigenfunctions are restricted by the fact that $d^j_{m,q}(\theta),\CD^{(j)}_{m,q}(\theta,\phi)$ are only well defined for $|m|,|q|\leq j$ and $m-j,q-j\in \IZ$. These restrictions imply that the allowed set of $j,m$ are given by 
\eq{
j&=q_{{R_A}}+\half+n_j\geq0\quad, \quad n_j \in \IZ~,\\
m&=-j,-j+1,...,j-1,j~.
}
where $q_{{R_A}}=max_{\mu \in \Delta_{R_A}}\big\{\langle \mu, T_3\rangle\big\}$. Note that when $n_j<0$ that we must impose additional restrictions on the $f_{a,\pm}$:
\eq{
f_{a,+}(r):=0~\text{ if } ~\left|q_a+\half\right|>j~,\\
f_{a,-}(r):=0~\text{ if }~\left|q_a-\half\right|>j~,
}
which give  $SU(2)_J$-representations  a nested structure. 

Because of this nested structure, it is useful to introduce additional notation. Let us define the spin-$j$ $R$-weight space $\Delta_R^{(j)}$ which is a restriction of the $R$-representation space to the subspace that contributes to the spin-$j$ representation of $\psi_R$:
\eq{
\Delta_R^{(j)}=\left\{\mu\in \Delta_R~\Big{|}~ |\langle \mu,T_3\rangle|\leq j+\half\right\}~.
}
Analogously, we can define the spin-$j$ restriction of the $SU(2)_T$-restricted $R_A$-weight spaces 
\eq{
\Delta_{R_A}^{(j)}=\left\{\mu\in \Delta_{R_A}~\Big{|}~|\langle \mu,T_3\rangle|\leq j+\half \right\}~.
}
For each $R_A$ and fixed $j$ we can also define an ordered basis $\{\mu_a\}$ of $\Delta_{R_A}^{(j)}$ such that $\langle \mu_a,T_3\rangle>\langle \mu_{a+1},T_3\rangle$ and $\langle \mu_a,\gamma_m\rangle\geq\langle \mu_{a+1},\gamma_m\rangle$. 
Notably, this defines a top and bottom weight $\mu_{top}=\mu_1$, $\mu_{bot}=\mu_{2j+1}$  with respect to $T_3$. 

The restriction to $\Delta_{R}^{(j)}$ allows us to define the basis of the spin-$j$ eigenfunctions of $\psi_R$ as 
\eq{
\big\{\text{spin-}j\big\}_R={\rm span}_\IC\big\{\CY^j_{m,q_a,s}\big\}\quad, \quad m=-j,...,j~,~s=\pm~,~\mu_a\in \Delta_R^{(j)}~,
}
which can be decomposed into irreducible $SU(2)_T$ representations as 
\eq{
\big\{\text{spin-}j\big\}_{R_A}={\rm span}_\IC\big\{\CY^j_{m,q_a,s}\big\}\quad, \quad m=-j,...,j~,~s=\pm~,~\mu_a\in \Delta_{R_A}^{(j)}~,
}
where we omit the $\CY^j_{m,q_a,s}$ for $\langle\mu_a,T_3\rangle=s\times \left(j+\half\right)$. Additionally, note that $ \Delta_{R_A}^{(j)}=\Delta_{R_A}$ if $j>q_{R_A}$.

We can also further decompose the $SU(2)_J$ representation into $\vS,\vK$ representations by computing the value of $2\vS\cdot \vK$ on the above $\psi_{R_A}$:
\eq{
(2\vS\cdot \vK)\,\psi_{R_A}&
=\left(\begin{array}{cc}
T_3-\half~&~-D_{T_3-1/2}^+\\
D_{T_3+1/2}^-~&~-T_3-\half
\end{array}\right)\,\psi_{R_A}~.
}
We see that acting on $\CY^j_{m,q_a}=\left(\begin{array}{c}f_{a}\,e^{i \phi/2}\,\CD^{(j)}_{-m,q_a+\half}(\theta,\phi)\\
g_{a}\,e^{-i \phi/2}\CD^{(j)}_{-m,q_a-\half}(\theta,\phi)
\end{array}\right)$, $2\vS\cdot \vK$ is given by:
\eq{
(2\vS \cdot\vK)\,\CY^j_{m,q_a}=\left(\begin{array}{cc}
q_a-\half&\sqrt{(j+1/2)^2-q_a^2}\\
\sqrt{(j+1/2)^2-q_a^2}&-q_a-\half
\end{array}\right)\CY^j_{m,q_a}
~,}
which has eigenvalues 
\eq{
{\rm Eig}[2\vS\cdot \vK]=\begin{cases}
j&j=k+\half\\
-j-1&j=k-\half
\end{cases}
} 
where $k$ is the spin of the $SU(2)_K$ representation which is generated by the $\vK$. 
This matches the standard equations for addition of angular momentum. This means that we can further decompose the representations 
\eq{\label{SdotKrep}
\CY^{j=k\pm \half}_{m,q_a}=\frac{1}{\sqrt{2\pi(2j+1+2q_a)}}\left(\begin{array}{c}
(j+1/2\pm q_a)\,e^{i \phi/2}\,\CD^{(j)}_{-m,q_a+\half}(\theta,\phi)\\
\pm \sqrt{(j+1/2)^2-q_a^2}\,e^{-i \phi/2}\CD^{(j)}_{-m,q_a-\half}(\theta,\phi)
\end{array}\right)~.
}

\subsection{Zero-Modes for General Spherically Symmetric Monopole}

Now we wish to solve the Dirac equation for a Weyl fermion in a representation $R$ of $SU(N)$ in the presence of a spherically symmetric monopole specified by $\{\vT,\vI\,\}$. The spherical symmetry of the monopole configuration implies that  the associated Dirac operator commutes with $\vJ$, and thus that we can expand the fermion zero-mode solutions in $SU(2)_J$ representations. 


Let us first restrict $R$ to a $SU(2)_T$-irreducible component $R_A$ and correspondingly take $\psi_{R_A}$ which we can expand with respect to a basis of the representation space 
\eq{
\psi_{R_A}=\sum_{a}\psi_{R_A}^a(r,\theta,\phi)\,v_{\mu_a}\quad, \quad \mu_a\in \Delta_{R_A}~.
} 
Now  consider the Dirac operator. Explicitly, it can be written
\eq{\label{Diracinit}
\bar\sigma^\mu D_\mu=&\sigma^{\hat{r}}\partial_r+\frac{1}{r}\left(\sigma^{\hat{\theta}}\partial_\theta+\frac{\sigma^{\hat\phi}}{\sin\theta}\Big((\partial_\phi+i T_3)-i T_3 \cos\theta\Big)\right)\\
&\quad-\half M_+e^{- i \phi}(\sigma^{\hat{\theta}}- i \sigma^{\hat{\phi}})+\half M_-e^{i\phi}(\sigma^{\hat{\theta}}+i \sigma^{\hat{\phi}})~,
}
where here we use the notation 
\begin{align}\begin{split}
&\sigma^{\hat{r}}=\sigma^3\cos\theta+(\sigma^1\cos\phi+\sigma^2\sin\phi)\sin\theta~,\\
&\sigma^{\hat{\theta}}=-\sigma^3\sin\theta+(\sigma^1\cos\phi+\sigma^2\sin\phi)\cos\theta~,\\
&\sigma^{\hat{\phi}}=-\sigma^1\sin\phi+\sigma^2\cos\phi~. 
\end{split}\end{align}
If we now perform a frame rotation
\be
\psi_{R_A,\alpha}=U_\alpha^{~\beta}\hat\psi_{R_A,\beta}\quad, \quad 
U^{~\beta}_{\alpha}=e^{\frac{i \phi}{2}\sigma^0\bar\sigma^3}e^{\frac{i\theta}{2}\sigma^0\bar\sigma^2}~,
\ee
the Dirac equation simplifies
\eq{
\Bigg[&\sigma^3\left(\partial_r+\frac{1}{r}\right)+\frac{\CK}{r}
-\left(\sigma^-e^{-i \phi}M_+-\sigma^+ e^{i \phi}M_-\right)\Bigg]\hat\psi_{R_A}=0~.
}
where
\eq{\label{ckdef}
\CK&=
\sigma^1\left(\partial_\theta+\half \cot(\theta)\right)+\frac{\sigma^2}{\sin\theta}\Big((\partial_\phi+i T_3)-i \cos(\theta)T_3\Big)=\left(\begin{array}{cc}
0&D^+_{T_3-\half}\\
D^-_{T_3+\half}&0\end{array}\right)~,
}
where again
\eq{
D^\pm_{q}=\partial_\theta\pm \frac{1}{\sin\theta}\Big(-i(\partial_\phi+i T_3)-q\cos\theta\Big)~. 
}

Now, we can restrict to a solution of fixed spin-$j=q_{R_A}+\half+n_j\geq0$ for $n_j\in \IZ$ which restricts  $\mu_a\in \Delta_{R_A}^{(j)}$. In the representation basis, the spin-$j$ representation can be written  
\eq{
\hat\psi_{R_A}=\sum_{\mu_a\in \Delta_{R_A}^{(j)}}\hat\psi_{R_A}^a v_{\mu_a}\quad, \quad \hat\psi_{R_A}^a = \left(\begin{array}{c}
f_{a}(r)\,e^{i \phi/2} \CD^{(j)}_{-m,q_{a}+\half}(\theta,\phi)\\
g_{a}(r)\,e^{-i \phi/2} \CD^{(j)}_{-m,q_{a}-\half}(\theta,\phi)
\end{array}\right)~,
}
where $f_{a}(r)=0$ for $|q_a+\half|>j$ and $g_{a}(r)=0$ for $|q_a-\half|>j$. 
With this ansatz, $\CK$ acts as
\eq{
\CK \cdot \hat\psi_{R_A}^a=\left(\begin{array}{cc}
0&-\sqrt{(j+1/2)^2-q_a^2}\\
\sqrt{(j+1/2)^2-q_a^2}&0
\end{array}\right)\hat\psi_{R_A}^a~.
}



Upon restricting to the spin-$j$ representation, 
the Dirac equation reduces
\eq{
\bar\sigma^\mu D_\mu\hat\psi_{R_A}=\frac{1}{r}\sigma^3 \left[r \partial_r+\sigma^3{\CK}+e^{-i \phi}\sigma^-M_++e^{i \phi}\sigma^+M_-+1\right]\hat\psi_{R_A}=0~.
}
This reduces to a first order matrix ODE for $f_a(r),g_a(r)$:
\eq{\label{diffeqF}
\left[r \partial_r+\CD\right]F=0\quad, \quad F=\sum_{\mu_a\in  \Delta_{R_A}}  \left(\begin{array}{c}
f_{a}(r)\\
g_{a}(r)
\end{array}\right)v_{\mu_a}~,
}
where 
\eq{
\CD=\sigma^3{\CK}\big{|}_j+ \sigma^-M_++ \sigma^+M_-+1~,
}
and $\sigma^3{\CK}\big{|}_j$ is a symmetric, constant matrix acting diagonally on the spin-$j$ restriction of  $\hat\psi_{R_A}$. Now, using the fact that ${M}_\pm(r)$ is only dependent on $r$ we find the solutions:
\eq{
F(r)=\CP\, \exp\left\{-\int_0^r \CD(r') \frac{dr'}{r'}\right\} F_0~, 
}
where $F_0$ is a constant vector and $\CP\, \exp$ is the path ordered exponential. Since $M_\pm(r)$ are smooth, bounded solutions, the smooth plane-wave normalizable solutions now reduces to finding the vectors $F_0$ that lead to solutions which fall off at least as fast as $1/r$ at $r\to \infty$ and are bounded  as $r\to 0$. 

Using the fact that 
\eq{
\lim_{r\to 0}M_\pm=T_\pm\quad, \quad \lim_{r\to \infty} M_\pm=I_\pm~, 
}
we can identify the asymptotics $\CD(r)$ as 
\eq{
\lim_{r\to 0}
\CD(r)=\CD_0:=\sigma^3\CK+\sigma^-T_++\sigma^+T_-+1
~,\\
\lim_{r\to \infty}
\CD(r)=\CD_\infty:=\sigma^3\CK+\sigma^-I_++\sigma^+I_-+1~.
}
We can then reexpress $\CD_0$ as 
\eq{
\CD_0=-2\vec{S}\cdot \vec{L}~.
}
Thus, solving for the regular solutions of the Dirac equation at $r\to 0$  reduces to solving for the decomposition of the solutions of $\psi_{R_A}$ into $\vL,\vS$ representations with non-negative $\vL\cdot \vS$. 

Similarly, in the limit $r\to \infty$, we find
\eq{
\CD_\infty=-2\vS\cdot (\vK-\vI)-2S_3(K_3-I_3)~.
}
Thus, solving for the plane-wave normalizable solutions of the Dirac equation at $r\to \infty$ reduces to solving for the decomposition into $\vS,(\vK-\vI)$ representations with non-positive $\vS\cdot(\vK-\vI)$. 

Finding the plane-wave normalizable solutions then reduces to computing the intersection between the non-negative eigenspaces of $\vL\cdot \vS$ and non-positive eigenspaces $\vS\cdot(\vK-\vI)$.

\subsubsection{Solutions of $r\to 0$}

Let us now study solutions of the Dirac equation in the limit $r\to 0$. Here we can solve for the spectrum of the solutions to the Dirac equation exactly. 

Let us  
restrict to the spin-$j= q_{R_A}+\half+n_j\geq0$ for $n_j\in \IZ$.  
Since $\CD_0=-\vS\cdot \vL$, we can compute the eigenvalues of $\CD_0$ by decomposing  the spin-$j$ $SU(2)_J$ representation as a tensor product of 
of $\vS$-, $\vL$-, and $\vT$- representations. 
We can then compute the eigenvalues of $2\vS\cdot \vL$ 
\eq{
{\rm Eig}[2\vS\cdot \vL]
=\begin{cases}
\ell&j\in q_{R_A}\otimes (\ell+\half)\\
-(\ell+1)&j\in q_{R_A}\otimes(\ell-\half)
\end{cases}
}
which correspond to solutions of the Dirac equation with the asymptotic behavior 
\eq{
\lim_{r\to 0}\psi_{R_A}\sim r^{2\vS\cdot \vL}\psi_0~.
}
In particular, the regular solutions at $r\to0$ are those for which $2\vS\cdot\vL\geq0$ which constrains:
\eq	{
\ell=\begin{cases}
n_j,n_j+1,...,2q_{R_A}+n_j&n_j\geq0\\
-n_j-1,-n_j,...,2q_{R_A}+n_j&n_j<0
\end{cases}
}

Now we can solve for the eigenvectors of $\CD_0$. First, recall that the space of spin-$j$ representations arises from the tensor product of $\vA=\vL+\vS$ and $\vT$-representations. 
In the vector gauge, the $A^2$ eigenfunctions with $A^2=(\ell+ 1/2)(\ell+3/2 )$ are of the form 
\eq{
\hat\psi^a_{R_A}=\left(\begin{array}{c}f_{a}(r)\,e^{i \frac{\phi}{2}}
\CD^{(\ell+\half)}_{-m_{a+},1/2}(\theta,\phi)\\
f_{a}(r)\,e^{-i \frac{\phi}{2}}
\CD^{(\ell+ \half)}_{-m_{a-},-1/2}(\theta,\phi)
\end{array}\right)~,
}
which we can see from \eqref{SdotKrep} by taking $q_a=0$. 

To relate this to the ``canonical gauge'' we have to gauge transform
\eq{
\hat\psi^a_{R_A}\to g\cdot\left(\begin{array}{c}f_{a}(r)\,
e^{i \frac{\phi}{2}}\CD^{(\ell+\half)}_{-m_{a+},1/2}(\theta,\phi)\\
f_{a}(r)\,
e^{-i \frac{\phi}{2}}\CD^{(\ell+ \half)}_{-m_{a-},-1/2}(\theta,\phi)
\end{array}\right)\quad, \quad g=e^{-i \phi T_3}e^{- i \theta T_2}e^{i \phi T_3}~.
}

Since $g$ is a gauge transformation, we can consider the spin up and spin down transformations separately: 
\eq{
\hat\psi_{{R_A}\pm}^a=e^{\pm i \frac{\phi}{2}}\sum_b\CD^{(q_{R_A})}_{q_{R_A}-a,q_{R_A}-b}(\theta,\phi)\left(\begin{array}{c}
f_{0}(r)~\CD^{(\ell+\half)}_{m_{ 0\pm},\pm \half}(\theta,\phi)\\
f_{1}(r)~\CD^{(\ell+\half)}_{m_{1\pm},\pm \half}(\theta,\phi)\\
f_{2}(r)~\CD^{(\ell+\half)}_{m_{2\pm},\pm \half}(\theta,\phi)\\
\vdots\\
f_{2q_{R_A}}(r)~\CD^{(\ell+\half)}_{m_{2q_{R_A}\pm},\pm \half}(\theta,\phi)
\end{array}\right)^b~.
}
Note here that the gauge transformation takes the form of a Wigner-$D$ matrix. This follows from the fact that the gauge transformation is a rotation in $SU(N)$ along the  embedding $SU(2)_T\hookrightarrow SU(N)$. 

Now using fact that Wigner $D$-functions multiply as 
\eq{
\CD^{(j_1)}_{m,m'}\CD^{(j_2)}_{n,n'}=\sum_{J,M,M'}\langle J,M|j_1,m;j_2,n\rangle \langle J,M'|j_1,m';j_2,n'\rangle \CD^{(J)}_{M,M'}~,
}
where  $\langle J,M|j_1,m_1;j_2,m_2\rangle$ are Clebsch-Gordon coefficients, 
we can determine the $\{f_{a}(r)\}$ that lead to a uniform spin-$j$ representation upon gauge transformation. To do so, first note that the transformation is only a   $J_3$-eigenfunction if
\eq{
m_{a\pm}=-m-(q_{R_A}-a)~. 
}

Now want to restrict to $J^2=j(j+1)$ eigenfunctions. 
For each $\ell=|j-q_{R_A}|-\half,|j-q_{R_A}|+\half,...,j+q_{R_A}-\half$, there exists a unique choice of $\{f_{a}(r)\}$ so that the resulting solution has purely spin-$j$ components. To find these solutions we must solve  
\eq{
&\left(\sum_{b=0}^{m+q_{R_A}-\ell-\half}\left\langle J,-m\big{|}\ell+\half,-m-q_{R_A}+b;q_{R_A},q_{R_A}-b\right\rangle f_{b}(r)\right)\\
&\qquad\times \left\langle J,q_{R_A}-a\pm \half \big{|}\ell+\half,\pm \half;q_{R_A},q_{R_A}-a\right\rangle=\begin{cases}
0&J\neq j\\
1&J=j
\end{cases}
}
Note that the $a$-dependence  factors out. 

For fixed $m,\ell,q_{R_A}$, the Clebsch-Gordon coefficients $\langle J,-m|\ell+\half,-m-q_{R_A}+a;q_{R_A},q_{R_A}-a\rangle$ define an orthogonal square matrix
\eq{
M_{J,a}^{(m,\ell)}=\langle J,-m|\ell+\half,-m-q_{R_A}+a;q_{R_A},q_{R_A}-a\rangle~.
}
Since each $M_{J,a}^{(m,\ell)}$ has an inverse,
 we can pick
\eq{
f_{b}(r)=f_m(r)\,(M^{(m,\ell)})^{-1}_{b,J}~\delta^{J,j}~. 
}
This leads to the solution
\eq{
\hat\psi^{(\ell)a}_{R_A}=f(r)\,\left(\begin{array}{c}
\langle j,q_{{R_A}}-a+\half|\ell+\half,\half;q_{R_A},q_{R_A}-a\rangle\, e^{i \frac{\phi}{2}}\CD^{(\ell+\half)}_{-m,q_{{R_A}}-a+1/2}(\theta,\phi)\\
\langle j,q_{{R_A}}-a-\half|\ell+\half,-\half;q_{R_A},q_{R_A}-a\rangle\, e^{-i \frac{\phi}{2}}
\CD^{(\ell+ \half)}_{-m,q_{{R_A}}-a-1/2}(\theta,\phi)
\end{array}\right)~,
}
which has the corresponding asymptotic behavior in the limit $r\to 0$:
\eq{
\lim_{r\to 0}f(r)\sim r^\ell+O(r^{\ell+\delta})\quad, \quad \delta>0~. 
}

We can reorganize the corresponding $F_0$ into the the vector 
\eq{
\hat{F}_0^{(\ell)}&:=v_{a+}^{(\ell)}\oplus v_{a-}^{(\ell)}~,\\
v_{a+}^{(\ell)}&=\left\langle j,q_{R_A}-a+\half\Big{|}\ell+\half,\half;q_{R_A},q_{R_A}-a\right\rangle~,\\
v_{a-}^{(\ell)}&=\left\langle j,q_{R_A}-a-\half\Big{|}\ell+\half,-\half;q_{R_A},q_{R_A}-a\right\rangle\quad, \quad  a=0,...,  2q_{R_A}~.
}


\subsubsection{Solutions at $r\to \infty$}

Now we want to solve for the normalizable solutions at $r\to \infty$. Here we are instead looking at the eigenfunctions of 
\eq{
\CD_\infty=-
2\vS\cdot (\vK-\vI)-2 S_3(K_3-I_3)~.
}

To do this, we can take a different approach. 
In the limit $r\to \infty$ there exists a gauge transformation 
\eq{
\omega=e^{i \phi I_3}e^{i \theta I_2}e^{-i \phi I_3}~,
}
which takes the asymptotic gauge field 
\eq{
A=2T_3 A_{Dirac}+\ihalf I_+ e^{-i \phi}(d\theta-i \sin\theta d\phi)-\ihalf I_- e^{i \phi}(d\theta+i \sin\theta d\phi)+O(1/r)~,
}
to the  abelian gauge 
\eq{
A_{ab.}=\omega^{-1} A\omega-i \omega^{-1}d \omega=2(T_3-I_3)A_{Dirac}+O(1/r)~. 
}
Since we are looking for the asymptotic solutions to the Dirac operator, it is sufficient to solve the Dirac equation in the abelian gauge at leading order in $1/r$. We are then reduced to solving the Dirac equation in the presence of an abelian monopole at long range, which has a simple, known solution. 

Now let us consider an irreducible component of the embedding of  $SU(2)_I\subset SU(N)$ of spin-$s_I$.  Further, let $R_{A,s_I}$ be the restriction of the $R_A$ representation to an irreducible spin-$s_I$ $SU(2)_I$ representation (which we will also  refer to by $s_I$) with corresponding weight system $\Delta_{R_{A,s_I}}$.

Consider the restriction of $\psi_{R_A}$ to $R_{A,s_I}$ and let us decompose
\eq{
\psi_{R_{A,s_I}}=\sum_{\mu_a\in \Delta_{R_{A,s_I}}}\psi_{R_{A,s_I}}^av_{\mu_a}~.
}
As before, the gauge transformation to the abelian gauge and frame rotation changes $\vJ$:
\eq{
\vJ\to&-i \hat{r}\times\Big(\vec\partial+i(T_3-I_3)\vA_{Dirac}-\ihalf \sigma^3\cot\theta\hat\phi\Big)\\&\quad
+\half I_+(\hat\theta+i \hat\phi)+\half I_-(\hat\theta-i \hat\phi)+\left(T_3+\frac{\sigma^3}{2}\right)\hat{r}~.
}

After performing the same frame rotation as before, we find that the Dirac equation here simplifies to 
\eq{
\left[r \partial_r+\widetilde\CD_\infty+1\right]F=0\quad, \quad F=\sum_a v_{\mu_a} \left(\begin{array}{c}
f_{a}(r)\\
g_{a}(r)
\end{array}\right)\quad, \quad \mu_a\in \Delta_{R_{A,s_I}}~,
}
where 
%
\eq{
\widetilde\CD_\infty= \omega^{-1}\CD_\infty\omega-1=\left(\begin{array}{cc}0&-D^+_{T_3-I_3-\half}\\
D^-_{T_3-I_3+\half}&0\end{array}\right)~,
}
which again allows us to decompose the asymptotic eigenfunctions of $\widetilde\CD_\infty$ in terms of  Wigner-$D$ functions. Explicitly, we can take the ansatz
\eq{
\hat\psi^a_{R_{A,s_I}}= \frac{1}{r}\left(\begin{array}{c}
{f}_a(r)\, e^{i \frac{\phi}{2}}\CD_{-m,Q_a+\frac{1}{2}}^{(n)}(\theta,\phi)\\
{g}_a(r)\,e^{-i \frac{\phi}{2}} \CD_{-m,Q_a-\frac{1}{2}}^{(n)}(\theta,\phi)\\
\end{array}\right)\quad, \quad \langle \gamma_m,\mu_a\rangle=Q_a~,
}
where $n$ is a fixed integer. Now, using the fact that $[\vI,\gamma_m]=0$ we have that 
\eq{
Q_a=\langle \mu_a\gamma_m\rangle=Q_{s_I}~,
}
is independent of $\mu_a\in \Delta_{R_{A,s_I}}$. 

Acting on $\hat\psi^a_{R_{A,s_I}}$ with quantum number $n$, we see that 
\eq{
\widetilde\CD_\infty\,\hat\psi^a_{R_{A,s_I}}=\left(\begin{array}{cc}
0&\sqrt{(n+1/2)^2-\frac{Q_{s_I}^2}{4}}\\
\sqrt{(n+1/2)^2-\frac{Q_{s_I}^2}{4}}&0
\end{array}\right)\hat\psi^a_{R_{A,s_I}}
}
which has eigenvalues $\pm \sqrt{(n+1/2)^2-\frac{Q_{s_I}^2}{4}}$ and corresponding eigenvectors $f_a=\pm g_a$. 

Now we wish to undo the gauge transformation to relate these solutions back to the spin-$j$ eigenfunctions in the canonical gauge. To do so, we must act by the gauge transformation $\omega^{-1}$ which is explicitly given by 
\eq{
\rho_{s_I}( \omega^{-1})_{ab}=\CD^{(s_I)}_{s_I-a,s_I-b}(\theta,\phi)\quad, \quad a,b=0,..., 2s_I+1~,
}
where $\CD^{(j)}_{-m,m'}(\theta,\phi)$ is the Wigner $D$-matrix and $\rho_{s_I}:SU(2)_I\to SU(N)$ is the embedding associated to the spin-$s_I$ representation. Due to the multiplication of Wigner $D$-functions, we find that given a fixed $s_I$, that the allowed values for $n$ that lead to a spin-$j$ representation are   $n=|j-s_I|,...,j+s_I$. 

After projecting onto the spin-$j$ representation, we find that the solutions of the asymptotic Dirac equation are of the form 
\eq{
\psi^a_{R_{A,s_I}}=U(\theta,\phi)f(r)\left(\begin{array}{c}
\langle j,q_a+1/2 |n,Q_{s_I}+1/2;s_I,s_I-a\rangle~e^{i \frac{\phi}{2}}\, \CD^{(j)}_{-m,q_a+1/2}(\theta,\phi)\\
\pm \langle j,q_a-1/2|n,Q_{s_I}-1/2;s_I,s_I-a\rangle~e^{-i \frac{\phi}{2}}\,\CD^{(j)}_{-m,q_a-1/2}(\theta,\phi)
\end{array}\right)~,
}
where 
$q_a=Q_{s_I}+s_I-a=\langle \mu_a,T_3\rangle$.

Thus, for each quantum number $n=|j-s_I|,...,j+s_I$ there are two linearly independent spin-$j$ eigenvectors of $\widetilde\CD_\infty$ 
with
  eigenvalues 
\eq{
\lambda_\pm^{(n)}=\pm\sqrt{(n+1/2)^2-\frac{Q_{s_I}^2}{4}}~.
}

This implies that the plane-wave normalizable modes at $r\to \infty$ are the solutions:
\eq{
\psi_{R_{A,s_I}}^{(n)}=U(\theta,\phi)f(r)\left(\begin{array}{c}
\langle j,q_a+1/2|n,Q_{s_I}+1/2;s_I,s_I-a\rangle~e^{i \frac{\phi}{2}}\CD^{(j)}_{-m,q_a+1/2} \\
\langle j,q_a-1/2|n,Q_{s_I}-1/2;s_I,s_I-a\rangle~e^{-i \frac{\phi}{2}}\CD^{(j)}_{-m,q_a-1/2}
\end{array}\right)~,
}
where $q_a=Q_{s_I}+s_I-a$ 
which have asymptotic fall-off
\eq{
\psi_{R_{A,s_I}}^{(n)i}\sim r^{-1-\sqrt{(n+1/2)^2-\frac{Q_{s_I}^2}{4}}}\psi_0~.
}
Note that as opposed to the eigenvectors of $\CD_0$ that are generically non-zero on the spin-$j$ restriction of $R_A$, the eigenvectors of $\widetilde\CD_\infty$ are generically non-zero on the restrictions of $R_A$ to the irreducible $\vI$-representations. Because of this decomposition, the only total angular momentum representations that come  from an irreducible  $\vI$ component of spin-$s_I$ are those of total spin-$j$ for $ j^{(s_I)}_{min}\leq j \leq j^{(s_I)}_{max}$ where
\eq{
j^{(s_I)}_{max}=max_{\mu \in \Delta_{R_A,s_I}}[\langle \mu,T_3\rangle]-\half\quad, \quad j^{(s_I)}_{min}=min_{\mu \in \Delta_{R_A,s_I}}[\langle \mu,T_3\rangle]-\half~.
}
Note that the only spin up/down solutions are those for which $n<\left|Q_{s_I}\mp \half\right|$ which corresponds to the choices 
\eq{
n=\begin{cases}
Q_{s_I}-\half & Q_{s_I}>0~\Rightarrow\text{  spin down}\\
-Q_{s_I}+\half & Q_{s_I}<0~\Rightarrow\text{ spin up}\\
\end{cases}
}
with no solution for $Q_{s_I}=0$. 

We can again identify this solution with the vector 
\eq{
\hat{F}_0^{(n)}&=u_{+,a}^{(n)}\oplus u_{-,a}^{(n)}~,\\
u_{\pm,a}^{(n)}&=\begin{cases}\pm\langle j,Q_{s_I}\pm1/2+s_I-a|n,Q_{s_I}\pm 1/2;s_I,s_I-a\rangle&\mu_a\in s_I\\
0&else
\end{cases}
}

\subsubsection{Normalizable Solutions}

Now let us go on to construct the plane-wave normalizable solutions of the zero-energy Dirac equation. This requires matching the solutions at $r\to \infty$ that fall off at least as fast as $1/r$ with the solutions at $r\to 0$ that are finite. 

Now we will show that any non-trivial, plane-wave normalizable, zero-energy solution of the Dirac equation falls off like $1/r$ at $r\to \infty$. Given a plane-wave normalizable zero-energy solution of the Dirac equation for $\psi_{R_A}$, we can construct the vector fields
\eq{
J_{R_A}^{~\,\mu} =\bar\psi_{R_A} \bar\sigma^\mu  \psi_{R_A}\quad, \quad \widetilde{J}_{R_A}^{~\,\mu} =\bar\psi_{R_A} \Phi\bar\sigma^\mu  \psi_{R_A}~,
}
where $\Phi$ is the Higgs field that makes up the monopole configuration. 
By assumption $\psi_{R_A}$ is smooth and goes to zero at $r\to \infty$ at least as fast as $1/r$ and $\Phi$ is smooth and has the asymptotic behavior 
\eq{
\lim_{r\to \infty}\Phi=\Phi_\infty-\frac{\gamma_m}{r}+O(1/r^{1+\delta})\quad, \quad\delta>0~.
}
Because of this, it is clear that 
\eq{
\int \partial_\mu   J^{~\,\mu}_{R_A}\,d^3x=\int_{S^2_\infty}J^{~\,r}_{R_A}\,d^2x<\infty\quad, \quad \int \partial_\mu   \widetilde{J}^{~\,\mu}_{R_A}  \,d^3x=\int_{S^2_\infty}\widetilde{J}^{~\,r}_{R_A}\,d^2x<\infty~. 
}
Further, if either of these quantities are non-zero, then it is clear that $\psi_{R_A}$ falls of as $1/r$ at $r\to \infty$.  

Using the fact that $\psi_{R_A}$ is a solution of the Dirac equation, we see that 
\eq{
\partial_\mu  J^{~\,\mu}_{R_A}=0\quad, \quad 
\partial_\mu  \widetilde{J}^{~\,\mu}_{R_A}
&=\bar\psi_{R_A}\bar\sigma^\mu (D_\mu \Phi) \psi_{R_A}~.
}
Thus, 
\eq{
\int \bar\psi_{R_A}\bar\sigma^\mu (D_\mu\Phi) \psi_{R_A}\,d^3x\neq 0\quad\Longrightarrow\quad \lim_{r\to \infty}r\psi_{R_A}\neq 0~.
}

Now, applying the Bogomolny equation:
\eq{
D_\mu\Phi=B_\mu~,
}
we can rewrite 
\eq{
\partial_\mu  \widetilde{J}^{~\,\mu}_{R_A}=\bar\psi_{R_A}\bar\sigma^\mu (D_\mu\Phi) \psi_{R_A}&=\bar\psi_{R_A}\bar\sigma^\mu  B_\mu  \psi_{R_A}=\bar\psi_{R_A}\bar\sigma^{\mu\nu} F_{\mu\nu} \psi_{R_A}~.
}
Using 
\eq{
[D_\mu,D_\nu]=F_{\mu\nu}\quad, \quad \sigma^{[\mu\nu]}=\sigma^\mu \bar\sigma^\nu+\bar\sigma^0 \delta^{\mu\nu}~,}
 and the fact that $\psi_{R_A}$ is annihilated by $\bar\sigma^\mu D_\mu$, we can further simplify:
\eq{
\partial_\mu \widetilde{J}^{~\,\mu}_{R_A}=\bar\psi_{R_A}\Big((\sigma^\mu  D_\mu )(\bar\sigma^\mu D_\mu)- \bar\sigma^0 D^2\Big)\psi_{R_A}
=-\bar\psi_{R_A}\bar\sigma^0D^2 \psi_{R_A}~.
}
Then, by integration by parts:
\eq{-
\int \bar\psi_{R_A}\bar\sigma^0D^2 \psi_{R_A}\,d^3x
&=-\int_{S^2_\infty}\bar\psi_{R_A} \bar\sigma^0D^r \psi_{R_A}\,d^2x+\int D_\mu\bar\psi_{R_A} \bar\sigma^0D^\mu \psi_{R_A}\,  d^3x\\
&=\int D_\mu\bar\psi_{R_A}\bar\sigma^0 D^\mu \psi_{R_A}\,  d^3x=\int |D_\mu \psi_{R_A}|^2d^3x >0~,
}
for any non-trivial $\psi_{R_A}$ solving the Dirac equation. Here we used the fact that 
\eq{
\int_{S^2_\infty}\bar\psi_{R_A} \bar\sigma^0D^r \psi_{R_A}\,d^2x=0~,
}
due to the fall off of at least $1/r$ of $\psi_{R_A}$ at $r\to \infty$. 

Therefore, for any $\psi_{R_A}\neq 0$ solving the massless Dirac equation in the presence of a spherically symmetric monopole
\eq{
\int \partial_\mu  \widetilde{J}^{~\,\mu}_{R_A}\,d^3x>0~.
}
Hence,  any such $\psi_{R_A}$  falls off like $1/r$ at $r\to \infty$.  

We can also use the  fact that $J^{~\,\mu}_{R_A}$ is divergenceless to determine further constraints on $\psi_{R_A}$. If we decompose 
\eq{
\psi_{{R_A}}=\sum_{\mu_a \in \Delta_{R_A}}\psi^a_{{R_A}} v_{\mu_a}\quad, \quad \psi^a_{{R_A}}=U(\theta,\phi)\left(\begin{array}{c}
f_a(r)\,e^{i \frac{\phi}{2}}\CD^{(j)}_{-m,q_a+\frac{1}{2}}(\theta,\phi)\\
g_a(r)\,e^{-i \frac{\phi}{2}}\CD^{(j)}_{-m,q_a-\frac{1}{2}}(\theta,\phi)
\end{array}\right)~,
}
then integrating the divergence of $J_{R_A}^{~\,\mu}$, implies that 
\eq{
\int \partial_\mu  J^{~\,\mu}_{R_A} d^3x=0\quad\Longrightarrow\quad
\label{rinftyconstraint}
\lim_{r\to \infty}r^2\sum_a\left(|f_a|^2-|g_a|^2\right)=0~. 
}

\section{Proof of Formula}
\label{sec:4}

Now we will prove the main result of this paper:

\thm~Given a fermion $\psi_R$ that transforms in the representation $R$ of $SU(N)$, the number of plane-wave normalizable solutions to the Dirac equation in the presence of a spherically symmetric monopole field configuration specified by the embeddings $\vT,\vI:\fsu(2)_{T,I}\hookrightarrow \fsu(N)$ is given by 
\eq{
\dim_{\IC}\big[\ker_R[\bar\sigma^\mu D_\mu]\big]=
\sum_{\mu \in \Delta_R}n_R(\mu)\langle \mu,T_3\rangle~sign\big(\langle \mu,T_3-I_3\rangle\big)~. 
}\\

To prove this we will need to make use of the following lemmas. In the following we will always assume that we decompose $\psi_R$ into irreducible $SU(2)_T$ representations and restrict to a single irreducible component $\psi_{R_A} $ which transforms in the spin-$q_{R_A}$ representation of $SU(2)_T$.

\lem{1} When $j> q_{R_A}$ there are no zero-eigenvalues of $\widetilde\CD_\infty$. \\

\pf Spin-$j$ zero-eigenvalues of $\widetilde\CD_\infty$ only occur if there exists fermions with weight $\mu\in \Delta_{R_{A,s_I}}$ that have quantum number $n=|\langle \mu,T_3-I_3\rangle|-\half>0.$ These only exist if $j+s_I\geq |\langle \mu,T_3-I_3\rangle|-\half\geq |j-s_I|$ where $v_\mu$  belongs to some irreducible $\vI^{(i,a)}$-component of spin-$s_I$. 
Now, using the fact that $|\langle \mu,T_3-I_3\rangle|\leq q_{R_A}-s_I,$ $\forall \mu \in \Delta_{R_{A,s_I}}$and the fact that $j\geq q_{R_A}+\half>s_I$ we see that 
\eq{
j-s_I\geq q_{R_A}-s_I+\half >| \langle \mu,T_3-I_3\rangle|-\half \quad, \quad \forall \mu \in \Delta_{R_{A,s_I}}~,
}
and 
therefore there are no zero-eigenvectors of $\widetilde\CD_\infty$ when $j>q_{R_A},s_I$. \qed

%




\lem{2} Pick $j<q_{R_A}$ and let $\mu_{top}$ ($\mu_{bot}$) be the top (resp. bottom) weight of $\Delta_{R_A}^{(j)}$ and let us take the convention $\langle \mu_{top},T_3\rangle>0$. There exists one single spin down zero-eigenvector of $\widetilde\CD_\infty$ with total spin-$j$ if  $\langle \mu_{top},T_3-I_3\rangle>0$ and one spin up zero-eigenvector of $\widetilde\CD_\infty$   with total spin-$j$ if $\langle \mu_{bot},T_3-I_3\rangle<0$. This means that the eigenvectors satisfy either: $f_a=0$ or $g_a=0$ respectively. There are no other spin-$j$ zero-eigenvectors of $\widetilde\CD_\infty$ when restricted to $R_A$.\\ 


\pf Let us fix $j<q_{R_A}$ which defines a highest and lowest weight $\mu_{top},\mu_{bot}\in  \Delta_{R_A}$. First, consider the highest weight component. In general, this belongs to a spin-$s_I$ representation of $SU(2)_I$. The fermions $\psi^a_{R_{A,s_I}}$ for $\mu_a \in \Delta^{(j)}_{R_{A,s_I}}$ are of the form 
\eq{\label{lemmatransform}
\psi_{R_{A,s_I}}^a=\sum_b \CD^{(s_I)}_{s_I-a,s_I-b}(\theta,\phi)\left(\begin{array}{c}
f_{b}(r)\,e^{i \frac{\phi}{2}}\CD^{(n)}_{-m-s_I+b,Q_{s_I}+\half}(\theta,\phi)\\
g_b(r)\,e^{-i\frac{\phi}{2}}\CD^{(n)}_{-m-s_I+b,Q_{s_I}-\half}(\theta,\phi)
\end{array}\right)~. 
}
Since the $\widetilde\CD_\infty$ eigenvalues of such a solution are $\lambda_\pm^{(n)}=\pm \sqrt{(n+1/2)^2-Q_{s_I}^2}$, it follows that there exists a  spin down zero-eigenvector of $\widetilde\CD_\infty$ with $n=Q_{s_I}-\half$ iff $Q_{s_I}-\half\geq0$ and it is the unique zero-eigenvector belonging to this $\vI$-representation. Conversely, if $Q_{s_I}\leq 0$, then there is no spin down zero-eigenvector from the irreducible $\vI$ representation. 

Thus, let us assume that $Q_{s_I}>0$. This condition combined with $q_{R_A}>j,s_I$ implies that 
\eq{\label{jrange}
Q_{s_I}-\half+s_I\geq j\geq Q_{s_I}-\half-s_I\geq0~.
}
Now, let us pick $n=Q_{s_I}-\half\geq0$. We now must show that the solutions constructed above have a non-trivial total spin-$j$ solution which requires showing that $[j]\in [n]\otimes [s_I]$. 

By using the standard multiplication of Wigner $D$-functions, we can expand the product of Wigner $D$-functions in \eqref{lemmatransform} as representations with total spin:
\eq{
\left[Q_{s_I}-\half\right]\otimes \left[s_I\right]=\left[Q_{s_I}-\half-s_I\right]\oplus \left[Q_{s_I}+\half-s_I\right]\oplus...\oplus\left[Q_{s_I}-\half+s_I\right]~,
}
where $[p]$ is the representation of total spin-$p$. It then follows from \eqref{jrange} that $[j]\in \left[Q_{s_I}-1/2\right]\otimes \left[s_I\right]$ and that there is a single spin down zero-eigenvector of $\widetilde\CD_\infty$ in the irreducible $\vI$ component containing $\mu_{top}$ when $\langle \mu_{top},\gamma_m\rangle>0$. 
The explicit solution is given by 
\eq{
f_a=0\quad, \quad g_a=\begin{cases}
 \left\langle j,Q_{s_I}+s_I-a-\half\big{|}Q_{s_I}-\half,Q_{s_I}-\half;s_I,s_I-a\right\rangle&\mu_a\in s_I\\
0&else
\end{cases}
}
where we have chosen a basis $\{\mu_a\}_{a=0}^{2s_I}$ of $\Delta_{R_{A,s_I}}$

Now, since the eigenvectors of $\widetilde\CD_\infty$ decompose with respect to irreducible $\vI$-representations, it follows that the unique spin down zero-eigenvector of $\psi_{R_A}$ with  total spin-$j$ of $\widetilde\CD_\infty$ is the one constructed above containing $\mu_{top}\in  \Delta_{R_A}$. 
The reason is that any spin-$j$ eigenvector $F_0^{(j)}$ of $\widetilde\CD_\infty$ that belongs to a different $\vI$-irreducible component of spin-$s_I'$ (i.e. the spin-$s_I'$ representation does not contain $v_{\mu_{top}}$) will not be spin polarized as we have shown in Section \ref{sec:fermrep} by explicit construction. This follows from the fact that $[\gamma_m,\vI]=0$ which, when restricted to an irreducible $SU(2)_T$ representation implies $Q_{s_I}\neq Q_{s_I'}$ even when $s_I=s_I'$. 

Then, since the gauge transformation $\omega$ that maps the abelian, asymptotic gauge to the canonical gauge commutes with $\vec{S}$, any such $F_0^{(j)}$ will necessarily correspond to a solution that is not polarized and hence with quantum number $n$ in the asymptotic gauge such that 
\eq{
n\neq \begin{cases}
Q_{s_I'}-\half & Q_{s_I'}>0\\
-Q_{s_I'}+\half&Q_{s_I'}<0
\end{cases}
}
Using the fact that $\lambda^{(n)}_\pm=\pm \sqrt{(n+1/2)^2-Q_{s_I'}^2}$,  we see that any such $F_0^{(j)}$ cannot be a spin-$j$ zero-eigenvector of $\CD_\infty$ and that the only spin-$j$ zero-energy solution is the one constructed above. 
Therefore, we see that there is only a single spin down zero-eigenvector of $\widetilde\CD_\infty$ with total spin-$j$ iff  $\langle \mu_{top},T_3-I_3\rangle>0$ coming from the irreducible $\vI$ component containing $\mu_{top}$ as constructed above. 

%

A nearly identical argument shows that there only exists a spin up zero-eigenvector of $\widetilde\CD_\infty$ when $\langle \mu_{bot},T_3-I_3\rangle<0$. \qed

\lem{3} 
If $\mu_{top},\mu_{bot}\in  \Delta_{R_A}$ are the highest/lowest weight of $\Delta_{R_A}^{(j)}$ for $j<q_{R_A}$ and 
\eq{
\langle \mu_{top},T_3-I_3\rangle\langle\mu_{bot},T_3-I_3\rangle=\langle \mu_{top},T_3-I_3\rangle\langle w(\mu_{top}),T_3-I_3\rangle<0~,
}  
then there exsits 2 zero-eigenvectors of $\widetilde\CD_\infty$ which are oppositely spin polarized.  Conversely, if 
\eq{
\langle \mu_{top},T_3-I_3\rangle\langle\mu_{bot},T_3-I_3\rangle=\langle \mu_{top},T_3-I_3\rangle\langle w(\mu_{top}),T_3-I_3\rangle\geq0~,
}
 then there is at most one zero-eigenvalue of $\widetilde\CD_\infty$
 where  $w$ is the unique element of the Weyl group of $SU(2)_T\subset SU(N)$. 
\\

\pf From Lemma 2, we know that if  $\langle \mu_{top},T_3-I_3\rangle>0$ and $\langle \mu_{bot},T_3-I_3\rangle<0$ then there are a pair of zero-eigenvalues of $\widetilde\CD_\infty$ that are oppositely spin polarized. 

Conversely, we see that if $\langle \mu_{top},T_3-I_3\rangle,\langle\mu_{bot},T_3-I_3\rangle\geq0$ or $\langle \mu_{top},T_3-I_3\rangle,\langle\mu_{bot},T_3-I_3\rangle\leq0$ there is at most a single spin polarized eigenvalue. 

Now, using the fact that $q_{R_A}>s_I$ and $[\vI,\gamma_m]=0$ for irreducible representations $R_A$, we see that if $\langle \mu_{top},T_3-I_3\rangle<0$ then $\langle \mu_{bot},T_3-I_3\rangle<0$ and similarly if $\langle \mu_{bot},T_3-I_3\rangle>0$ then $\langle \mu_{top},T_3\rangle>0$. This implies that if 
\eq{
\langle \mu_{top},T_3-I_3\rangle\langle\mu_{bot},T_3-I_3\rangle<0~,
}
then $\langle \mu_{top},T_3-I_3\rangle>0$ and $\langle \mu_{bot},T_3-I_3\rangle<0$ and $\widetilde\CD_\infty$ has two spin-$j$ zero-eigenvectors. 

Further, since we are considering $R_A$ as a representation of $SU(2)_T$, the unique non-trivial element $w$ of the Weyl group of $\fsu(2)_T$ exchanges the top and bottom weight so that 
\eq{
\mu_{bot}=w(\mu_{top})~. 
}
Explicitly, in terms of Dynkin indices for weights:
\eq{
\mu=[m_1,...,m_{N-1}]~~\Rightarrow ~~w(\mu)=[-m_{N-1},...,-m_1]~. }
Therefore, there exists two zero-eigenvectors of $\widetilde\CD_\infty$ if 
\eq{
\langle \mu_{top},T_3-I_3\rangle\langle\mu_{bot},T_3-I_3\rangle=\langle \mu_{top},T_3-I_3\rangle\langle w(\mu_{top}),T_3-I_3\rangle<0~.}

\qed

\lem{4} When there are no zero-eigenvalues of $\widetilde\CD_\infty$, there are no zero-energy plane-wave normalizable solutions of the Dirac equation.\\

\pf  First let us recall that a zero-eigenvalue of $\widetilde\CD_\infty$ corresponds to a solution of the Dirac equation with asymptotic behavior 
\eq{
\lim_{r\to \infty}\psi_{R_A}\sim \frac{1}{r}\psi_{R_A}^{(0)}~. 
}
And therefore, if there are no zero-eigenvalues of $\widetilde\CD_\infty$, then 
\eq{
\lim_{r\to \infty}r\psi_{R_A}=0~.
}

Now using the fact that for any non-trivial solution of the Dirac equation $\psi_{R_A}$:
\eq{
\int \partial_\mu \widetilde{J}_{R_A}^{~\,\mu}\,d^3x=\int |D_\mu \psi_{R_A}|^2\,d^3 x>0\quad, \quad \widetilde{J}^{~\,\mu}_{R_A}=\bar\psi_{R_A}\bar\sigma^a \Phi\psi_{R_A}~,
} we see that 
\eq{
\lim_{r\to \infty}\int_{S^2} \widetilde{J}^r_{R_A}\, \,r^2\,d^2\Omega\neq0~~\Rightarrow~~\lim_{r\to \infty}r \,\psi_{R_A}(r)\neq 0~. 
}
Thus, any solution of the Dirac equation $\psi_{R_A}$ must have leading order $1/r$ as $r\to \infty$. Therefore, if $\widetilde\CD_\infty$ has no zero-eigenvalues, then all asymptotic solutions that fall off faster than $1/r$ and there can be no solution to the Dirac equation.
\qed

\lem{5} When $\widetilde\CD_\infty$ has a single pair of zero-eigenvalues, there is at most a single, plane-wave normalizable,  zero-energy solution to the Dirac equation. Conversely, if $\widetilde\CD_\infty$ has  less than 2 zero-modes, then there is no normalizable solution of the Dirac equation. \\

\pf From Lemma 4, we see that any non-trivial solutions to the Dirac equation has 
\eq{
\lim_{r\to \infty}r\psi_R\neq 0~. 
}
From our explicit solutions to the Dirac equation, we see that these solutions correspond to the zero-eigenvectors of $\widetilde\CD_\infty$. From Lemma 2, we know that there are at most two such eigenvectors and that they are polarized. 

Then, for any time-independent solution of the Dirac equation $\psi_{R_A}$:
\eq{
\partial_\mu  J^{~\,\mu}_{R_A}=0\quad, \quad J_{R_A}^{~\,\mu}=\bar\psi_{R_A}\bar\sigma^\mu\psi_{R_A}~.
}
This implies 
\eq{\label{polcond}
0=\int \partial_\mu  J^{~\,\mu}_{R_A}\,d^3x=\lim_{r\to \infty}\int J^r_{R_A}\,r^2\,d^2\Omega=\lim_{r\to \infty}r^2\sum_a\left(|f_a|^2-|g_a|^2\right)~.
}
This constraint then implies that the two polarized zero-eigenvectors of $\widetilde\CD_\infty$ can at most combine into one non-trivial solution of the Dirac equation. 

Further, if there is only a single zero-eigenvalue of $\widetilde\CD_\infty$, then from Lemma 2 it must be spin polarized. The condition \eqref{polcond} then cannot be satisfied except by the trivial solution. Additionally, if there are no zero-eigenvalues of $\widetilde\CD_\infty$, then there are no solutions that go like $1/r$ at $r\to \infty$ and hence from Lemma 4, we see that there can be no zero energy plane-wave normalizable solutions of the Dirac equation. 
\qed

\subsection{Proof of Formula}

First let us decompose $\psi_R$ into irreducible representations of $SU(2)_T\subset SU(N)$:
\eq{
\psi_R=\sum_A \psi_{{R_A}}\quad, \quad R=\bigoplus_A {R_A}~. 
}
From Lemma 4 and 5, we know that there is at most a single solution to the Dirac equation if there are a pair of zero-eigenvalues of $\widetilde\CD_\infty$ and further from Lemmas 1,2, and 3 that these only occur for spin-$j$ representations where $j<q_{R_A}$ and 
\eq{
\langle \mu_{top},T_3-I_3\rangle\langle w(\mu_{top}),T_3-I_3\rangle<0~,
}
where $w$ is the element of the Weyl group of $SU(2)_T\subset SU(N)$.  
It then remains to show that this contribution is indeed non-trivial for all such $j$.

Let us recall that the spin-$j$ solutions of the Dirac equation when restricted to the $SU(2)_T$-irreducible representation $R_A$ are of the form 
\eq{
\psi_{R_A}=U(\theta,\phi)\sum_{\mu_a\in  \Delta_{R_A}}\left(\begin{array}{c}
f_{a}(r)\,e^{i \phi/2} \CD^{(j)}_{-m,q_{a}+\half}(\theta,\phi)\\
g_{a}(r)\,e^{-i \phi/2} \CD^{(j)}_{-m,q_{a}-\half}(\theta,\phi)
\end{array}\right)v_{\mu_a}~,
}
where 
\eq{
&F(r)=\sum_{\mu_a\in  \Delta_{R_A}}\left(\begin{array}{c}
f_a(r)\\
g_a(r)
\end{array}\right)v_{\mu_a}\quad, \quad F(r)=\CP\,e^{-\int_0^r \CD(r')\frac{dr'}{r'}}F_0~, \\&
\CD(r)=\sigma^3 \CK\big{|}_j+\sigma^-M_++\sigma^+M_-+1~.
}
and 
 $\CK\big{|}_j$ is the differential operator $\CK$ in \eqref{ckdef} acting on the spin-$j$ representation. 

Any constant vector $F_0$ that leads to a vector $F(r)$ such that 
\eq{
I.)~\lim_{r\to \infty}r F(r)<\infty\quad, \quad II.)~\lim_{r\to 0}F(r)<\infty~,
}
is smooth and plane-wave normalizable. This follows from the fact that $\CD(r)$ is smooth and  bounded for all finite $r$ which is due to the fact that we are considering a smooth monopole configuration and $\CK\big{|}_j$ is a constant. 

To any such $F_0$, we can identify a $\hat{F}_0$ which is simply a reorganization into spin up plus spin down components:
\eq{
\hat{F}(r)&=\left(\begin{array}{c}f_1(r)\\\vdots\\f_{2q_{R_A}+1}(r)\\g_1(r)\\ \vdots\\g_{2q_{R_A}+1}(r)\end{array}\right)=f_a(r)\oplus g_a(r)=\CP\,e^{-\int_0^r \hat\CD(r')\frac{dr'}{r'}}\hat{F}_0~.
}
Above, we have solved explicitly for the set of $\hat{F}_0$ that correspond to the set of $F_0$ which provide solutions that obey condition $I.)$ and that obey condition $II.)$ separately. Explicitly, they are given by 
\eq{
I.)~
\hat{F}_0^{(\ell)}&:=v_{a+}^{(\ell)}\oplus v_{a-}^{(\ell)}~,\\
v_{a+}^{(\ell)}&=\left\langle j,q_{R_A}-a+\half\Big{|}\ell+\half,\half;q_{R_A},q_{R_A}-a\right\rangle~,\\
v_{a-}^{(\ell)}&=\left\langle j,q_{R_A}-a-\half\Big{|}\ell+\half,-\half;q_{R_A},q_{R_A}-a\right\rangle\quad, \quad  a=1,...,  2q_{R_A}~,\\
II.)~
\hat{F}_0^{(n)}&=u_{+,a}^{(n)}\oplus u_{-,a}^{(n)}~,\\
u_{\pm,a}^{(n)}&=\begin{cases}\pm\langle j,Q_{s_I}\pm1/2+s_I-a|n,Q_{s_I}\pm 1/2;s_I,s_I-a\rangle&\mu_a\in \Delta_{R_{A,s_I}}\\
0&else
\end{cases}
}
These can be truncated to form $4j=2j+2j$-dimensional vectors. 

Since the allowed $\ell=q_{R_A}-j-\half,...,q_{R_A}+j-\half$, there are $2j$ allowed $\hat{F}_0^{(\ell)}$. To determine the number of allowed eigenvectors $\hat{F}_0^{(n)}$ notice that  the eigenvalues always come in pairs $\lambda^{(n)}_\pm=\pm \lambda^{(n)}$. This, coupled with the fact that there are two zero-eigenvectors for $j<q_{R_A}$ and  $\langle \mu_{top},\gamma_m\rangle\langle w(\mu_{top}),\gamma_m\rangle<0$ implies that there are $2j+1$ allowed eigenvectors of $\hat{F}_0^{(n)}$. 

Now, using the fact that the $\{\hat{F}_0^{(\ell)}\}$ and $\{\hat{F}_0^{(n)}\}$ each form an orthogonal set of $4j$-dimensional vectors, we find that the set of plane-wave normalizable solutions to the Dirac equation is given by the intersection of the $2j$-dimensional and $(2j+1)$-dimensional hyperplanes in $\IR^{4j}$ spanned by the $\hat{F}_0^{(\ell)}$ and $\hat{F}_0^{(n)}$ respectively. It is now clear that the intersection is guaranteed to be at least a 1-dimensional space by dimension counting. Therefore, there always exists at least one spin-$j$ solution to the Dirac equation for each $\psi_{R_A}$ when $j<q_{R_A}$ and $\langle \mu_{top},\gamma_m\rangle\langle w(\mu_{top}),\gamma_m\rangle<0$ for $\mu_{top}\in \Delta_{R_A}^{(j)}$. 

Then, using the fact that the lemmas imply that there exists at most one spin-$j$ solution to the Dirac equation for each $\psi_{R_A}$ when $j<q_{R_A}$ and  $\langle \mu_{top},\gamma_m\rangle\langle w(\mu_{top}),\gamma_m\rangle<0$ for $\mu_{top}\in \Delta_{R_A}^{(j)}$, we see that under such conditions there is a  unique spin-$j$ solution to the Dirac equation for $\psi_{R_A}$.

We then find that the kernel of the Dirac operator decomposes as:
\eq{
\ker_{{R_A}}[\bar\sigma^\mu D_\mu]=\bigoplus_{\mu\in \Delta^+_{R_A}[T_3,I_3]}\left[\langle \mu,T_3\rangle-\half\right]~,
}
where $[j]$ is the total angular momentum representation of spin-$j$ and 
\eq{
\Delta^+_{R_A}[T_3,I_3]=\left\{\mu \in \Delta_{{R_A}}^+~\Big{|}~\langle \mu,T_3-I_3\rangle\langle w(\mu),T_3-I_3\rangle<0\right\}~,
}
and $w(\mu)$ is the image of $\mu$ with respect to the non-trivial element of the  Weyl group of $\fsu(2)_T\subset \fsu(N)$ and $\Delta^+_{R_A}$ is the set of positive weights of $\Delta_{R_A}$ with respect to $T_3$. 
It then follows that the dimension of the kernel is given by
\eq{
\dim_\IC\big[\ker_{{R_A}}[\bar\sigma^\mu D_\mu]\big]&=\sum_{\mu \in \Delta^+_{R_A}[T_3,I_3]}2\langle \mu,T_3\rangle\\
&= \lim_{\epsilon\to 0^+} \sum_{\mu \in \Delta_{R_A}}n_{R_A}(\mu)\langle \mu,T_3\rangle\, \sign\left(\langle \mu-\epsilon w(\mu), T_3- I_3\rangle\right)~,
}
where here the limit accounts for the cases when $\langle \mu,T_3-I_3\rangle\langle w(\mu),T_3-I_3\rangle= 0$. 
Then by adding all of the contributions from all of the representations $R=\bigoplus_A{R_A}$, we find 
\eq{
\dim_{\IC}\big[\ker_R[\bar\sigma^\mu D_\mu]\big]&=\lim_{\epsilon\to 0^+}\sum_A\sum_{\mu \in \Delta_{R_A}}\langle \mu,T_3\rangle\, \sign\left(\langle \mu-\epsilon w(\mu), T_3- I_3\rangle\right)\\
&= \lim_{\epsilon\to 0^+}\sum_{\mu \in \Delta_R}n_R(\mu)\langle \mu,T_3\rangle\, \sign\left(\langle \mu-\epsilon w(\mu), T_3- I_3\rangle\right)~.
}
\qed

\corro ~The plane-wave normalizable kernel of the Dirac operator in the presence of a spherically symmetric monopole decomposes into total angular momentum representations as 
\eq{
\ker_R[i\bar\sigma^\mu D_\mu]=\bigoplus_{\mu \in \Delta^+_R[T_3,I_3]}\left[\langle \mu,T_3\rangle-\half\right]^{\oplus n_R(\mu)}~,
}
where $[j]$ is the spin-$j$ representation. \\

\pf From the above analysis, we see that the kernel of the Dirac operator when restricted to the $R_A$-irreducible representation of $SU(2)_T$ decomposes into total angular momentum representations as 
\eq{
\ker_{{R_A}}[\bar\sigma^\mu D_\mu]=\bigoplus_{\mu\in \Delta^+_{R_A}[T_3,I_3]}\left[\langle \mu,T_3\rangle-\half\right]~,
}
where $[j]$ is the spin-$j$ representation. Summing over the irreducible components $R_A$, we find 
\eq{
\ker_R[\bar\sigma^\mu D_\mu]&=\bigoplus_A\bigoplus_{\mu\in \Delta^+_{R_A}[T_3,I_3]}\left[\langle \mu,T_3\rangle-\half\right]=\bigoplus_{\mu \in \Delta^+_R[T_3,I_3]}\left[\langle \mu,T_3\rangle-\half\right]^{\oplus n_R(\mu)}~.
}
\qed
 
 \section{Example: $SU(5)$ Monopoles with a Fundamental Fermion}
 \label{sec:5}
 
 Here we will illustrate the above index theorem with a simple example. Let us consider $SU(5)$ gauge theory with a Weyl fermion $\psi_{R_f}$ in the fundamental representation $R_f$. We would like to compute the number of fermion zero-energy solutions in the presence of a spherically symmetric monopole and determine their structure. Let us pick a spherically symmetric monopole with $\vT$ embedding specified by:
 \eq{
R_f( T_3)=\diag(2,1,0,-1,-2)~.
 }
Let us further specify a monopole with asymptotic magnetic charge 
 \eq{
R_f( \gamma_m)=\half\diag(1,1,1,1,-4)\quad\Longleftrightarrow\quad R_f( I_3)=\half\diag(3,1,-1,-3,0)~. 
 }
 
Using the fact that the weight multiplicity $n_f(\mu)=1$ for the fundamental representation, we find that the number of zero-energy solutions is given by 
\eq{
\dim_\IC\big[\ker_{{R_f}}[\bar\sigma^\mu D_\mu]\big]&= \sum_{\mu \in \Delta_{{R_f}}}\langle \mu,T_3\rangle~sign\big(\langle \mu,T_3-I_3\rangle\big)\\
&=(2\times 1)+(1\times 1)+0+(-1\times 1)+(-2\times -1)=4~,
}
which decomposes in terms of total angular momentum representations as 
\eq{
\ker_{R_f}[i\bar\sigma^\mu D_\mu]=\bigoplus_{\mu \in \Delta^+_{R_f}[T_3,I_3]}\left[\langle \mu,T_3\rangle-\half\right]^{\oplus n_f(\mu)}=\left[2-\half\right]=\left[\frac{3}{2}\right]~.
}

\subsection{Spin-1/2 Solution}

Let us first show that there is no spin-1/2 zero-energy solution. First let us introduce a basis of the representation and weight space 
\eq{
R_f={\rm span}_\IC\{v_a\}_{a=1}^5~,~\{\mu_a\}_{a=1}^5\in \Delta_{R_f}\quad, \quad R_f(T_3)v_a= \langle T_3,\mu_a\rangle v_a=(3-a)v_a 
~,
}
and similarly expand 
\eq{
\psi_{R_f}=\sum_{a=1}^5 \psi_av_a~.
}
By examining the eigenvalues of $T_3$, we then see that only $\psi_{2,3,4}$ can contribute to the spin-$1/2$ representation and further that they are of the form 
\eq{
\psi_2&=U(\theta,\phi)\left(\begin{array}{c}
0\\
g_{2}(r)\,e^{-i \phi/2} \CD^{(1/2)}_{-m,\half}(\theta,\phi)
\end{array}\right)~,\\
\psi_3&=U(\theta,\phi)\left(\begin{array}{c}
f_{3}(r)\,e^{i \phi/2} \CD^{(1/2)}_{-m,\half}(\theta,\phi)\\
g_{3}(r)\,e^{-i \phi/2} \CD^{(1/2)}_{-m,-\half}(\theta,\phi)
\end{array}\right)~,\\
\psi_4 &=U(\theta,\phi)\left(\begin{array}{c}
f_{4}(r)\,e^{i \phi/2} \CD^{(1/2)}_{-m,-\half}(\theta,\phi)\\
0
\end{array}\right)
}
where $F(r)=\Big(g_2(r),f_3(r),g_3(r),f_4(r)\Big)$ is given by
\eq{
F(r)=\CP\,e^{-\int_0^r \CD(r')\frac{dr'}{r'}}F_0\quad, \quad \CD(r)=\left(\begin{array}{cccc}
1&\alpha_2(r)&0&0\\
\alpha_2(r)&1&-1&0\\
0&-1&1&\alpha_3(r)\\
0&0&\alpha_3(r)&1
\end{array}\right)
}
where 
\eq{&
\lim_{r\to 0}\alpha_2(r)=\sqrt{6}\quad,\quad \lim_{r\to \infty}\alpha_2(r)=2~,\\&
\lim_{r\to 0}\alpha_3(r)=\sqrt{6}\quad,\quad \lim_{r\to \infty}\alpha_3(r)=\sqrt{3}~.
}
We then find that the asymptotic eigenvalues of $\widetilde{\CD}_\infty=\CD_\infty-1$ are given by 
\eq{
{\rm Eigenvalues}[\widetilde\CD_\infty]=\{\pm \sqrt{6},\pm \sqrt{2}\}~,
}
and hence by Lemma 4 there are no zero-energy fermion solutions. 

\subsection{Spin-3/2 Solution}

Now let us construct the spin-3/2 solution. Here, all of the $\psi_a$ contribute. Let us expand 
\eq{
\psi_a=U(\theta,\phi)\left(\begin{array}{c}
f_{a}(r)\,e^{i \phi/2} \CD^{(3/2)}_{-m,3-a+\half}(\theta,\phi)\\
g_{a}(r)\,e^{-i \phi/2} \CD^{(3/2)}_{-m,3-a-\half}(\theta,\phi)
\end{array}\right)~,
}
where $f_1(r)=g_5(r)=0$. 

Again the solution $F(r)=\Big(g_1(r),f_2(r),g_2(r),f_3(r),g_3(r),f_4(r),g_4(r),f_5(r)\Big)$ is given by 
\eq{
F(r)=\CP\,e^{-\int_0^r \CD(r')\frac{dr'}{r'}}F_0~,
}
where 
\eq{
\CD=\left(\begin{array}{cccccccc}
1&\alpha_1(r)&&&&&&\\
\alpha_1(r)&1&-\sqrt{3}&&&&&\\
&-\sqrt{3}&1&\alpha_2(r)&&&&\\
&&\alpha_2(r)&1&-2&&&\\
&&&-2&1&\alpha_3(r)&&\\
&&&&\alpha_3(r)&1&\sqrt{3}&\\
&&&&&-\sqrt{3}&1&\alpha_4(r)\\
&&&&&&\alpha_4(r)&1
\end{array}\right)
}
Here the $\alpha_I(r)$ have the limiting behavior:
\eq{
&\lim_{r\to 0}\alpha_1(r)=2\quad
\quad, \quad \lim_{r\to \infty}\alpha_1(r)=\sqrt{3}
~,\\
&\lim_{r\to 0}\alpha_2(r)=\sqrt{6}\quad\,,\quad \lim_{r\to \infty}\alpha_2(r)=2~,\\&
\lim_{r\to 0}\alpha_3(r)=\sqrt{6}\quad\,,\quad \lim_{r\to \infty}\alpha_3(r)=\sqrt{3}~,\\
&\lim_{r\to 0}\alpha_4(r)=2\quad
\quad, \quad \lim_{r\to \infty}\alpha_4(r)=0~.
}
Consequently, the eigenvalues of $\CD_0,\widetilde\CD_\infty$ are given by 
\eq{
{\rm Eigenvalues}[\widetilde\CD_\infty]&=\{0,0,\pm \sqrt{2},\pm \sqrt{6},\pm \sqrt{12}\}~,\\
{\rm Eigenvalues}[\CD_0]&=\{0,-1,-2,-3~,~2,3,4,5\}~.
}
 By computing the overlap of the non-negative eigenvectors of $\CD_0$ with the non-positive eigenvectors of $\widetilde\CD_\infty$, we find that the solution for $F_0$ is given numerically by 
 \eq{
 F_0&\approx v_{\ell=0}+0.55v_{\ell=1}+0.29v_{\ell=2}+0.10 v_{\ell=3}\\
 &\approx (-0.54,0.16,-0.62,0.27,-0.76,0.45,-1.24,1.58)~.
 }
 Further, the limiting behavior of $F(r)$ is given by 
 \eq{
 \lim_{r\to 0}F(r)=v_{\ell=0}+O(r)\quad, \quad \lim_{r\to \infty}F(r)=-1.58v_{top}+1.58v_{bot}+O(1/r^{\sqrt{2}})
 }
 where 
 \eq{
 v_{top}=\half(1,0,1,0,1,0,1,0)\quad, \quad v_{bot}=(0,0,0,0,0,0,0,1)~,
\\
v_{\ell=0}=\left(
\sqrt{\frac{4}{5}},
-\sqrt{\frac{1}{5}},
\sqrt{\frac{3}{5}},
-\sqrt{\frac{2}{5}},
\sqrt{\frac{2}{5}},
-\sqrt{\frac{3}{5}},
\frac{1}{\sqrt{5}},
-\sqrt{\frac{4}{5}}\right)~.
 }
 Note that $v_{top},v_{bot}$ are polarized and satisfy \eqref{rinftyconstraint} and that $v_{\ell=0}$ is an exact reordering of the Clebsch-Gordon coefficients
 \eq{
\left\langle \frac{3}{2},2-a\pm \half~\Big{|}\half,\pm \half;2,2-a\right\rangle~.
 }

\section*{Acknowledgements}

We would like to thank G. Satishchandran and J. Harvey for discussions and  A.B. Royston, G. Satishchandran, and M. Stern for comments on the draft. TDB is supported by the  Mafalda and Reinhard Oehme Postdoctoral Fellowship in the Enrico Fermi Institute at the University of Chicago and in part by DOE grant DE-SC0009924. 

\bibliographystyle{utphys}
\bibliography{FermionIndexBib}

\providecommand{\href}[2]{#2}\begingroup\raggedright\begin{thebibliography}{10}

\bibitem{Callias:1977kg}
C.~Callias, ``{Index Theorems on Open Spaces},''
  \href{http://dx.doi.org/10.1007/BF01202525}{{\em Commun. Math. Phys.}
  {\bfseries 62} (1978) 213--234}.

\bibitem{Jackiw:1975fn}
R.~Jackiw and C.~Rebbi, ``{Solitons with Fermion Number 1/2},''
  \href{http://dx.doi.org/10.1103/PhysRevD.13.3398}{{\em Phys. Rev. D}
  {\bfseries 13} (1976) 3398--3409}.

\bibitem{Gauntlett:1993sh}
J.~P. Gauntlett, ``{Low-energy dynamics of N=2 supersymmetric monopoles},''
  \href{http://dx.doi.org/10.1016/0550-3213(94)90457-X}{{\em Nucl. Phys. B}
  {\bfseries 411} (1994) 443--460},
  \href{http://arxiv.org/abs/hep-th/9305068}{{\ttfamily arXiv:hep-th/9305068}}.

\bibitem{Sethi:1995zm}
S.~Sethi, M.~Stern, and E.~Zaslow, ``{Monopole and Dyon bound states in N=2
  supersymmetric Yang-Mills theories},''
  \href{http://dx.doi.org/10.1016/0550-3213(95)00517-X}{{\em Nucl. Phys. B}
  {\bfseries 457} (1995) 484--512},
  \href{http://arxiv.org/abs/hep-th/9508117}{{\ttfamily arXiv:hep-th/9508117}}.

\bibitem{Gauntlett:1999vc}
J.~P. Gauntlett, N.~Kim, J.~Park, and P.~Yi, ``{Monopole dynamics and BPS dyons
  N=2 superYang-Mills theories},''
  \href{http://dx.doi.org/10.1103/PhysRevD.61.125012}{{\em Phys. Rev. D}
  {\bfseries 61} (2000) 125012},
  \href{http://arxiv.org/abs/hep-th/9912082}{{\ttfamily arXiv:hep-th/9912082}}.

\bibitem{Gauntlett:2000ks}
J.~P. Gauntlett, C.-j. Kim, K.-M. Lee, and P.~Yi, ``{General low-energy
  dynamics of supersymmetric monopoles},''
  \href{http://dx.doi.org/10.1103/PhysRevD.63.065020}{{\em Phys. Rev. D}
  {\bfseries 63} (2001) 065020},
  \href{http://arxiv.org/abs/hep-th/0008031}{{\ttfamily arXiv:hep-th/0008031}}.

\bibitem{Brennan:2016znk}
T.~D. Brennan and G.~W. Moore, ``{A note on the semiclassical formulation of
  BPS states in four-dimensional $N=$ 2 theories},''
  \href{http://dx.doi.org/10.1093/ptep/ptw159}{{\em PTEP} {\bfseries 2016}
  no.~12, (2016) 12C110}, \href{http://arxiv.org/abs/1610.00697}{{\ttfamily
  arXiv:1610.00697 [hep-th]}}.

\bibitem{Brennan:2018ura}
T.~D. Brennan, G.~W. Moore, and A.~B. Royston, ``{Wall Crossing from Dirac
  Zeromodes},'' \href{http://dx.doi.org/10.1007/JHEP09(2018)038}{{\em JHEP}
  {\bfseries 09} (2018) 038}, \href{http://arxiv.org/abs/1805.08783}{{\ttfamily
  arXiv:1805.08783 [hep-th]}}.

\bibitem{Georgi:1974sy}
H.~Georgi and S.~L. Glashow, ``{Unity of All Elementary Particle Forces},''
  \href{http://dx.doi.org/10.1103/PhysRevLett.32.438}{{\em Phys. Rev. Lett.}
  {\bfseries 32} (1974) 438--441}.

\bibitem{Callan:1982ah}
C.~G. Callan, Jr., ``{Disappearing Dyons},''
  \href{http://dx.doi.org/10.1103/PhysRevD.25.2141}{{\em Phys. Rev. D}
  {\bfseries 25} (1982) 2141}.

\bibitem{Callan:1982au}
C.~G. Callan, Jr., ``{Dyon-Fermion Dynamics},''
  \href{http://dx.doi.org/10.1103/PhysRevD.26.2058}{{\em Phys. Rev. D}
  {\bfseries 26} (1982) 2058--2068}.

\bibitem{Callan:1982ac}
C.~G. Callan, Jr., ``{Monopole Catalysis of Baryon Decay},''
  \href{http://dx.doi.org/10.1016/0550-3213(83)90677-6}{{\em Nucl. Phys. B}
  {\bfseries 212} (1983) 391--400}.

\bibitem{Rubakov:1982fp}
V.~A. Rubakov, ``{Adler-Bell-Jackiw Anomaly and Fermion Number Breaking in the
  Presence of a Magnetic Monopole},''
  \href{http://dx.doi.org/10.1016/0550-3213(82)90034-7}{{\em Nucl. Phys. B}
  {\bfseries 203} (1982) 311--348}.

\bibitem{Wilkinson:1977yq}
D.~Wilkinson and A.~S. Goldhaber, ``{Spherically Symmetric Monopoles},''
  \href{http://dx.doi.org/10.1103/PhysRevD.16.1221}{{\em Phys. Rev. D}
  {\bfseries 16} (1977) 1221}.

\bibitem{Brennan:2021ewu}
T.~D. Brennan, ``{Callan-Rubakov Effect and Higher Charge Monopoles},''
  \href{http://arxiv.org/abs/2109.11207}{{\ttfamily arXiv:2109.11207
  [hep-th]}}.

\bibitem{Murray:2003}
M.~K. Murray and M.~A. Singer, ``A note on monopole moduli spaces,''
  \href{http://dx.doi.org/10.1063/1.1590056}{{\em Journal of Mathematical
  Physics} {\bfseries 44} no.~8, (Aug, 2003) 3517–3531}.
  \url{http://dx.doi.org/10.1063/1.1590056}.

\bibitem{Bais:1997qy}
F.~A. Bais and B.~J. Schroers, ``{Quantization of monopoles with nonAbelian
  magnetic charge},''
  \href{http://dx.doi.org/10.1016/S0550-3213(97)00778-5}{{\em Nucl. Phys. B}
  {\bfseries 512} (1998) 250--294},
  \href{http://arxiv.org/abs/hep-th/9708004}{{\ttfamily arXiv:hep-th/9708004}}.

\bibitem{tHooft:1974kcl}
G.~'t~Hooft, ``{Magnetic Monopoles in Unified Gauge Theories},''
  \href{http://dx.doi.org/10.1016/0550-3213(74)90486-6}{{\em Nucl. Phys. B}
  {\bfseries 79} (1974) 276--284}.

\bibitem{Polyakov:1974ek}
A.~M. Polyakov, ``{Particle Spectrum in the Quantum Field Theory},'' {\em JETP
  Lett.} {\bfseries 20} (1974) 194--195.

\bibitem{Prasad:1975kr}
M.~K. Prasad and C.~M. Sommerfield, ``{An Exact Classical Solution for the 't
  Hooft Monopole and the Julia-Zee Dyon},''
  \href{http://dx.doi.org/10.1103/PhysRevLett.35.760}{{\em Phys. Rev. Lett.}
  {\bfseries 35} (1975) 760--762}.

\bibitem{Weinberg:2006rq}
E.~J. Weinberg and P.~Yi, ``{Magnetic Monopole Dynamics, Supersymmetry, and
  Duality},'' \href{http://dx.doi.org/10.1016/j.physrep.2006.11.002}{{\em Phys.
  Rept.} {\bfseries 438} (2007) 65--236},
  \href{http://arxiv.org/abs/hep-th/0609055}{{\ttfamily arXiv:hep-th/0609055}}.

\bibitem{Wilkinson:1978zh}
D.~Wilkinson and F.~A. Bais, ``{Exact SU(N) Monopole Solutions with Spherical
  Symmetry},'' \href{http://dx.doi.org/10.1103/PhysRevD.19.2410}{{\em Phys.
  Rev. D} {\bfseries 19} (1979) 2410}.

\end{thebibliography}\endgroup

\end{document}